\documentclass[11pt,a4paper]{article}

\usepackage[no-natbib-sort]{my-jheppub}

\usepackage[english]{babel}
\usepackage{amsthm}
\usepackage{graphicx} 

\newcommand{\besubeqs}{\begin{subequations}}
\newcommand{\esubeqs}{\end{subequations}}

\graphicspath{ {pics/} }

\newcommand{\cO}{{\cal O}}

\def \be {\begin{eqnarray}}
\def \ee {\end{eqnarray}}
\def\bea{\begin{eqnarray}}
\def\eea{\end{eqnarray}}
\def\nn{\nonumber}
\newcommand{\w}[1]{\\[0.#1cm]}

\newcommand{\pl}{\partial}

\begin{document}

\date{\currenttime}

%%%%%%%%%%%%%%%%%%%%%%%%%%%%%%%%%%%%%%%%%%%%%%%%%%%%%%%%%%%%%%%%%%%%%%%%%%%%%%%%%%%%%%%%%%%%
\title{On one loop corrections in higher spin gravity  }
%to a four-point function \\

\emailAdd{ponomarev@tamu.edu}
\emailAdd{sezgin@physics.tamu.edu}
\emailAdd{evgeny.skvortsov@aei.mpg.de}

\author[a,b]{Dmitry Ponomarev}
\author[a]{\quad Ergin Sezgin}
\author[c,b]{\quad Evgeny Skvortsov}

\affiliation[a]{Texas A\&M University, 
College Station, TX 77843, USA}
\affiliation[b]{Lebedev Institute of Physics, 
Leninsky ave. 53, 119991 Moscow, Russia}
\affiliation[c]{Albert Einstein Institute, Am Muhlenberg 1, D-14476, Potsdam-Golm, Germany}

\abstract{We propose an approach to compute one-loop corrections to the four-point amplitude in the higher spin gravities that are holographically dual to free $O(N)$, $U(N)$ and $USp(N)$ vector models. We compute the double-particle cut of one-loop diagrams by expressing them in terms of tree level four-point amplitudes. We then discuss how the remaining contributions to the complete one-loop diagram can be computed. With certain assumptions we find nontrivial evidence for the shift in the identification of the bulk coupling constant and $1/N$ in accordance with the previously established result for the vacuum energy. 
%
% More specifically, the one-loop correction to the four-point amplitude vanishes for the $U(N)$ model, and leads to % %the shift in the inverse gravitational coupling constant given by $N\to N-1$ for the $O(N)$ model and $N\to N+1$ for %the $USp(N)$ case. 
}

\unitlength = 1mm

%\today
%                                                                                                                                                          \date {}
\begin{flushright}\small{MI-TH-1919}\end{flushright}

\maketitle

%\def \lR {L}

%\def \ll {{\cal \ell}}

%\def \be {\begin{equation}}\def \ee {\end{equation}}

%\def \ads {AdS$_4$\ }
%\def \iffa {\iffalse} 

%\def \ed {\bibliography{KHSA}{}
%\bibliographystyle{utphys} \end{document}}

%%%%%%%%%%%%%%%%%%%%%%%%%%%%%%%%%%%%%%%%%%%%%%%%%%%%%%%%%%
\section{Introduction}
%%%%%%%%%%%%%%%%%%%%%%%%%%%%%%%%%%%%%%%%%%%%%%%%%%%%%%%%%%
%
Higher spin gravities (HSGRA) are theories that extend gravity with massless fields of spin greater than two. Presence of an infinite tower of gauge fields gives rise to infinite-dimensional symmetries which, in turn, constrain physical observables in these theories. It turns out that already these general symmetry considerations are  powerful enough to fix the observables  almost uniquely. These arguments also suggest that symmetries render HSGRA renormalizable, if not finite. Hence, one expects that HSGRA provides examples of  very economic  extensions of gravity, which can be consistently quantized, thereby providing simplest models of quantum gravity. The latter point of view is further supported by AdS/CFT correspondence, in which some of the HSGRA's are conjectured to be dual to rather simple conformal fields theories \cite{Sundborg:2000wp, Sezgin:2002rt,Klebanov:2002ja,Sezgin:2003pt,Gaberdiel:2010pz}. Holography also allows to infer that generic HSGRA's posses a number of unusual properties. In particular, it can be shown that higher spin interactions are nonlocal in a conventional sense \cite{Bekaert:2015tva,Sleight:2017pcz,Ponomarev:2017qab}. Moreover, unlike string theory, there is no limit in which the effective field theory description contains finitely many fields with suppressed higher derivative terms. This makes generic HSGRA hard to study even at the classical level. It is conceivable that a number of difficulties encountered in HSGRA treated as a field theory are akin to those which arise in string field theory, which calls for methods similar to those used in string field theory to tackle these problems. We shall comment on this similarity later. 

Despite the above mentioned generic difficulties, at present, there is a handful of examples of higher spin gravities  that avoid the nonlocality problem one way or another: (i) in three dimensions a class of HSGRA's are obtained as  generalizations of the Chern-Simons formulation of  gravity  \cite{Blencowe:1988gj,Campoleoni:2010zq,Henneaux:2010xg}; (ii) in any even dimension there is a well-defined higher spin extension of conformal gravity \cite{Tseytlin:2002gz,Segal:2002gd,Bekaert:2010ky}; (iii) lastly, there exists a chiral (or self-dual) higher spin theory in four dimensional flat and $AdS$ spaces that is formulated in the light-cone gauge \cite{Ponomarev:2016lrm,Ponomarev:2017nrr,Skvortsov:2018jea,Skvortsov:2018uru}. All these theories admit relatively simple actions, which enables their quantization. It is in these examples that quantization of HSGRA has been thoroughly studied \cite{Joung:2015eny,Ponomarev:2016jqk,Beccaria:2016syk,Hikida:2017ehf,Skvortsov:2018jea}.  Much less is known about quantum effects in the HSGRA's that have free CFT duals. The main obstacle is the lack of an action, except for the first few terms \cite{Bekaert:2014cea,Bekaert:2015tva,Sleight:2016dba,Skvortsov:2018uru} and these terms are not sufficient to carry out any complete one-loop computation. 

The only quantum effects in anti-de Sitter higher spin gravities that have been well-explored up to date require a free action: these are based on the computation of the one-loop free energy corrections started in \cite{Giombi:2013fka} and elaborated further in \cite{Giombi:2014iua,Giombi:2014yra,Beccaria:2014xda,Beccaria:2014jxa,Beccaria:2014zma,Beccaria:2014qea,Basile:2014wua,Beccaria:2015vaa,Beccaria:2016tqy,Bae:2016rgm,Bae:2016hfy,Pang:2016ofv,Gunaydin:2016amv,Giombi:2016pvg,Bae:2016xmv,Brust:2016xif}. These results are very important for the present paper. The free energy of a higher spin gravity admits an expansion in the coupling constant $G$ which is of order $1/N$. The leading term is determined by the value of the action in (Euclidean) AdS background, which is unknown.  Nevertheless, one can proceed to the one-loop correction, which requires only the knowledge of the spectrum and not of interactions. The spectrum is read off from the CFT dual. It turns out that the one-loop free energy does not always vanish, but it is proportional to the leading term expected from holography. This implies that the duality can work provided that the naive relation $G^{-1}\sim N$ is modified to $G^{-1}\sim (N+b)$, where $b$ is an integer. It has been shown that $b=-1,0,1$ for the free $O(N),\, U(N)$ and $USp(N)$ ($N$ even) vector models for all dimensions $d$ \cite{Giombi:2013fka}, including the fractional ones \cite{Giombi:2013fka,Skvortsov:2017ldz}. 

To extend these results to higher-point functions one can consider various paths. One of them is to use the Vasiliev equations \cite{Vasiliev:1992av}, which are generating equations that should give higher spin equations (differential equations for space-time fields) upon solving for the auxiliary variables. For various difficulties arising in this process, see \cite{Giombi:2009wh,Giombi:2010vg,Boulanger:2015ova,Skvortsov:2015lja,Vasiliev:2016xui,Gelfond:2017wrh,Misuna:2017bjb,Didenko:2018fgx}. At any rate, the existing results for vertices are not sufficient to carry out a full-fledged loop computation. Moreover, quantization of non-Lagrangian systems presents an additional problem \cite{Kazinski:2005eb}. While a formal action from which an extended version of the equations follow has been proposed \cite{Boulanger:2015kfa}, it has a non-standard form and its quantization and use for computations of amplitudes remains to be an open problem. 

Given the  obstacles  briefly alluded to above with regard to the quantization of higher spin gravity, we are led to consider an alternative approach in which we exploit a remarkably simple nature of their expected holographic duals. In this approach, we adopt a principle that the {\it classical} bulk HSGRA is defined by means of the holographic reconstruction starting from a well defined boundary conformal field theory \cite{Petkou:2003zz,Bekaert:2014cea,Bekaert:2015tva,Sleight:2016dba}.\footnote{There is a proposal for the all-order reconstruction \cite{Koch:2010cy} which employs an appropriate change of variables
% that is done via a clever change of variables 
in the free CFT path integral.} Once holographically reconstructed action is available, one can use the standard field theory techniques to compute loop diagrams, see e. g. \cite{Penedones:2010ue,Fitzpatrick:2011hu,Fitzpatrick:2011dm,Cardona:2017tsw,Giombi:2017hpr,Yuan:2017vgp,Bertan:2018khc,Bertan:2018afl} for recent loop computations in AdS.
However, an explicit holographic reconstruction of the classical bulk action is an arduous task, and has been carried out for lower order vertices only. The currently available part of the bulk higher spin action is sufficient to obtain only partial results for the one-loop self-energy diagram \cite{Giombi:2017hpr}. Extending holographic reconstruction to higher orders in interactions is conceptually straightforward, but increasingly complicated.
Another subtlety related to this analysis is that holographic reconstruction allows us to reproduce the action only up to on-shell trivial vertices or, equivalently, up to a freedom of field redefinitions. In the perturbative analysis such ambiguities normally do not have any impact on physical observables. However, in HSGRA's we are dealing with an infinite set of fields and infinite expansions in derivatives, and, as a result, field redefinitions should be treated with caution. For these reasons, quantization of HSGRA's based on the direct application of holographic reconstruction turns out to be very challenging.

Instead of applying the holographically reconstructed action for computations of loop diagrams directly  as outlined above, it is more promising to combine this approach with the on-shell methods. Namely, by generalizing to AdS the standard flat space techniques based on unitarity, analyticity and known high-energy behavior of amplitudes, one can compute the singular part of any loop diagram using the on-shell data for lower-loop and lower-point amplitudes and then reconstruct the complete amplitude up to certain local terms, using the AdS version of dispersion relations. Various forms of AdS on-shell methods have been developed and successfully applied for loop computations recently \cite{Fitzpatrick:2011hu,Fitzpatrick:2011dm,Alday:2016njk,Aharony:2016dwx,Caron-Huot:2017vep,Alday:2017xua,Aprile:2017bgs,Aprile:2017xsp,Alday:2017vkk,Aprile:2017qoy,Alday:2018pdi,Ghosh:2018bgd}. These techniques are particularly well-suited for one-loop computations in HSGRA's defined holographically, as they allow to bypass reconstruction of the bulk action, instead expressing the loop amplitude in terms of tree-level amplitudes, which, it turn, are readily given by correlators in the free CFT on the boundary.
In this paper, we shall use a version of on-shell methods to compute the double cut of the one-loop four-point diagram for scalar fields in HSGRA's dual to the free $O(N)$, $U(N)$ and $USp(N)$ vector models.  We then comment on how the complete one-loop amplitude can be reproduced.

At first sight one can claim that higher spin symmetries are unique and completely fix the holographic correlation functions \cite{Maldacena:2011jn,Boulanger:2013zza,Alba:2013yda,Alba:2015upa} and hence the problem of quantization is already solved. However, it is not guaranteed that, given an action, there exists a consistent quantization of this action and all the ambiguities (regularization, sums over infinitely many fields running in loops) can  consistently be resolved. The most important piece of information that is not provided by the CFT side at all is the nontrivial dependence $G(N)$ of the bulk coupling constant $G$ on the number of fields $N$ in the CFT dual. Our approach allows us to infer part of this information from the CFT.

The double-cut allows us to reconstruct a part of the full one-loop four-point amplitude. To obtain the missing single-particle contributions, we assume that the one-loop corrections to three-point amplitudes are consistent with the dependence $G$ on $N$ implied by vacuum one-loop corrections. There is a non-trivial test of the duality at this stage --- the total one-loop correction to the four-point amplitude has to agree with the assumed $G(N)$. Remarkably, we find our results to be in agreement with the computation of the free energy, discussed above: the one-loop correction to the four-point function in the theory dual to the free $U(N)$-model does vanish, while it is proportional to the tree-level result for the theories dual to the free $O(N)$ and $USp(N)$ models. In particular, we find $G^{-1}\sim N$ for the $U(N)$-case,  $G^{-1}\sim N-1$ for the $O(N)$-case and $G^{-1}\sim N+1$ for the $USp(N)$-case. There is also a puzzle left, regarding the disconnected contribution that we discuss at the very end.

The outline of the paper is as follows. In section \ref{sec:FK} we review how to compute the singular part or, the cut, of one-loop diagrams following \cite{Fitzpatrick:2011dm}. Some very basic facts about $U(N)$, $O(N)$ and $USp(N)$ models together with the remarks on the locality issues are given in section \ref{sec:freeCFT}. The cut is computed in section \ref{sec4}. In section \ref{sec:towards} we interpret our results and discuss the computation of the full one-loop amplitude. Conclusions and future directions can be found in section  \ref{sec:Discussion}.

%%%%%%%%%%%%%%%%%%%%%%%%%%%%%%%%%%%%%%%%%%%%%%%%%%%%%%%%%%
\section{Singularities in one loop from tree amplitudes}
\label{sec:FK}
%%%%%%%%%%%%%%%%%%%%%%%%%%%%%%%%%%%%%%%%%%%%%%%%%%%%%%%%%%

Focusing on the singularities of the one loop amplitudes, a time honored approach to study them uses unitarity which relates them to phase space integrals of products of lower-loop amplitudes. This approach works very well in Minkowski space and recently it has also been extended to AdS space. Here we shall closely follow the approach detailed in \cite{Fitzpatrick:2011dm} in  a toy model describing the quartic interactions of three distinct scalar fields. We shall first review briefly this approach. Then, we shall extend it for application to the higher spin gravities we are considering. As we shall see, this involves a number of new features and subtleties that are absent in the toy model of \cite{Fitzpatrick:2011dm}.

The model studied in \cite{Fitzpatrick:2011dm} has three scalars $\phi$, $\chi$ and $\psi$  in the bulk, which have different dimensions $\Delta_\phi$, $\Delta_\chi$ and $\Delta_\psi$ and their interactions are given by
\be
{\cal L}= \frac14 \lambda \phi^2\chi^2 + \frac14 g \chi^2\psi^2\ .
\label{int}
\ee
Consider the four-point one-loop Witten diagram with $\phi,\phi,\psi,\psi$ on external lines and $\chi$ running in the loop. The idea of \cite{Fitzpatrick:2011dm} is to express the singular part of this amplitude in terms of conformal block coefficients for tree-level diagrams  $\phi \phi \to \chi\chi$ and $\chi\chi\to \psi\psi$. 

The tree-level bulk diagram generated by the contact interaction $\lambda \phi^2\chi^2$ has the conformal block decomposition 
\begin{equation}
A_4^{{\rm tree}}(u,v) = \sum_{n,s}\delta a^{\phi\phi,\chi\chi}_{ [\phi\phi]_{n,s}} G^{\phi\phi,\chi\chi}_{[\phi\phi]_{n,s}}(u,v) + \delta a^{\phi\phi,\chi\chi}_{ [\chi\chi]_{n,s}} G^{\phi\phi,\chi\chi}_{[\chi\chi]_{n,s}}(u,v) \,
\label{30may3}
\end{equation}
where
\begin{equation}
\label{30may4}
\delta a^{\phi\phi,\chi\chi}_{ [\phi\phi]_{n,s}} = \bar c_{\phi\phi[\phi\phi]_{n,s}}\delta c_{\chi\chi[\phi\phi]_{n,s}}, \qquad  \delta a^{\phi\phi,\chi\chi}_{ [\chi\chi]_{n,s}}=
\delta c_{\phi\phi[\chi\chi]_{n,s}}\bar c_{\chi\chi[\chi\chi]_{n,s}}\ .
\end{equation}
Here $\bar c$ and $\delta c$ can be interpreted as the OPE coefficients at zeroth and first order in $\lambda$, $G(u,v)$ is the conformal block, with the superscript referring to the operators associated with external lines and the subscript referring to the operator exchanged in the conformal block. Furthermore $[\cO\cO]_{n,s}$ refers to double-trace operators of dimension $2\Delta_\cO+2n+s$ and spin $s$.  Note that being zeroth order in coupling constant, $\bar c$ refers to the OPE coefficients in the free theory, and ${\bar c}^2$ gives the conformal block coefficient for the disconnected diagrams. Similar formulae to those given above hold for the $\chi\chi\to \psi\psi$ amplitude. 

It was shown in \cite{Fitzpatrick:2011dm} that the contributions associated with the conformal blocks $G_{[\chi\chi]_{n,l}}$ for the one-loop bubble amplitude $\phi\phi\to\psi \psi$ with $\chi$ running in the loop are given by
\begin{equation}
{\rm disc}\left[A_4^{{\rm 1-loop}}(u,v)\right] = \sum_{n,s} \delta c_{ \phi\phi[\chi\chi]_{n,s}}\delta c_{\psi\psi[\chi\chi]_{n,s}}  G^{\phi\phi,\psi\psi}_{[\chi\chi]_{n,s}} (u,v).
\label{30may5}
\end{equation}
This part of the amplitude is expressed exclusively in terms of the on-shell data for tree-level amplitudes and in this respect is analogous to the singular part of flat space amplitudes computed using unitarity \footnote{{There are different representations for AdS amplitudes/ CFT correlators. In each of them there is a mathematically rigorous definition of what is meant by the singular part. In particular, when the correlator is given in the coordinate representation, its singular part is defined by the correlator's double discontinuity, which is a certain linear combination of the Euclidean correlator and its analytic continuations around branch cuts to the Lorentzian regime, see (2.14) of \cite{Caron-Huot:2017vep}. It can then be shown that the double discontinuity in the case we consider is insensitive to conformal blocks with $[\phi\phi]$ and $[\psi\psi]$ exchanged. The remaining contributions --- namely, $[\chi\chi]$ --- are then regarded as singular.}}. Moreover, it has been shown that in the flat space limit it reduces to the imaginary part of the flat space amplitude or, equivalently, to its discontinuity across the double-particle branch cut. This explains our notation on the left hand side of (\ref{30may5}).

 In the following it will be convenient to apply \eqref{30may5} in the form
\begin{equation}
{\rm disc}\left[A_4^{{\rm 1-loop}}(u,v)\right] = \sum_{n,s} \frac{\delta a^{\phi\phi,\chi\chi}_{[\chi\chi]_{n,s}}\ \delta a^{\chi\chi,\psi\psi}_{[\chi\chi]_{n,s}}}{\bar a_{[\chi\chi]_{n,s}}} \  G^{\phi\phi,\psi\psi}_{[\chi,\chi]_{n,s}} (u,v)\ ,
\label{30may6}
\end{equation}
where 
\begin{equation}
\label{30may2}
\bar a_{ [\chi\chi]_{n,s}} = \left(\bar c_{\chi \chi [\chi\chi]_{n,s}}\right)^2\ .
\end{equation}
With new features and subtleties that arise in the case of HSGRA taken into account, the formula  \eqref{30may6} will be the starting point for our calculation of the singular part of the one-loop amplitude arising from the double cut.

%%%%%%%%%%%%%%%%%%%%%%%%%%%%%%%%%%%%%%%%%%%%%%%%%%%%%%%%%%
\section{Some basics of free CFT's }
\label{sec:freeCFT}
%%%%%%%%%%%%%%%%%%%%%%%%%%%%%%%%%%%%%%%%%%%%%%%%%%%%%%%%%%
Generic CFT duals of anti-de Sitter higher spin gravities are free theories. In this paper we consider the simplest case --- the free vector models. There are several options that are almost indistinguishable as free theories, but result in significant effects in the bulk. We shall focus on those with $O(N)$, $U(N)$ and $USp(N)$ symmetry. Since the computations in free CFT's are trivial, in this section, we mainly introduce notation and list some of the operators that will be used later. As a byproduct, we will also see readily why higher spin gravities are not conventional field theories due to bulk nonlocalities.

\paragraph{The $\boldsymbol{O(N)}$, $\boldsymbol{U(N)}$ and $\boldsymbol{USp(N)}$ models.} 
In the $O(N)$ vector model the fundamental fields are $N$ real scalars $\phi^a(x)$ in the vector representation of $O(N)$. The fundamental fields in the $U(N)$ model are $N$ complex scalars $\bar{\phi}^a$ in the fundamental representation of $U(N)$ and the fundamental fields of the $USp(N)$ model are  $4N$ pseudo-real scalars $\phi^{A,a}$ where $A=1,...,2N$ ($N$ is even) and $a=1,2$, taking values in the bi-fundamental representation of $USp(N)\times USp(2)$.
We will choose the normalization of the two-point functions as follows:
\bea
O(N) &:&\qquad \langle \phi^a(x) \phi^b(0)\rangle =\frac{\delta^{ab}}{|x|^{d-2}}\equiv \delta^{ab} G(x)\ ,
\nn\\
U(N) &:&\qquad   \langle \bar{\phi}^a(x) \phi_b(0)\rangle =\frac{\delta^a_b}{|x|^{d-2}}\ ,
\nn\\
USp(N) &:& \qquad \langle \phi^{aA}(x) \phi^{bB}(0)\rangle =\frac{\epsilon^{ab} \Omega^{AB}}{|x|^{d-2}}\ ,
\eea
where $\epsilon^{ab}$ and $\Omega^{AB}$ are symplectic invariant tensors. The simplest single trace operators are
\bea
O(N) &:&\qquad J_0=\phi^a\phi^b\delta_{ab}\ ,
\nn\\
U(N) &:&\qquad J_0=\bar\phi^a\phi_a\ ,
\nn\\
USp(N) &:&\qquad J_0=\phi^{aA}\phi^{bB}\epsilon_{ab}\Omega_{AB}\ .
\eea
They have spin zero and conformal dimension $\Delta=d-2$. Their two-point and four-point functions can be obtained straightforwardly by Wick contractions, and take the form
\be
\langle J_0(x) J_0(0)\rangle = \frac{c_1}{(x^2)^{d-2}}\ ,
\ee
and
\bea
\langle J_0(x_1) J_0(x_2) J_0(x_3) J_0(x_4)\rangle &=&  c_2 (G_{12}^2G_{34}^2+G_{13}^2G_{24}^2+G_{14}^2G_{23}^2) +
\label{fpk1}\\
&& +c_3 \left(G_{12}G_{23}G_{34}G_{41}+G_{13}G_{34}G_{42}G_{21}+G_{14}G_{42}G_{23}G_{31}
\right)\ ,
\nn
\eea
where $G_{ij} := 1/|x_i-x_j|^{(d-2)}$ and
\be
(c_1,c_2,c_3) = \begin{cases}  (2N,4N^2,16N) & \mbox{for} \ O(N)\ ,\\ (N,N^2,2N) &\mbox{for} \  U(N)\ ,\\(8N,64N^2,64N) &\mbox{for} \  USp(N) \ .\end{cases}
\ee
The remaining single-trace operators are given in Appendix A. In the case of $O(N)$ they contain all even spins, each occurring once; in the case of $U(N)$ they contain all integer spins each occurring once, and in the case of $USp(N)$ they contain all even spins each occurring once, and all odd spins each occurring in a triplet of $USp(2)$. In the latter case, an example is given by the triplet of spin-one currents
\begin{align}
    J^{ab}_\mu&= \phi^{A,a} \overleftrightarrow{{}_{\phantom{\,\,}}\pl_\mu} \phi^{B,b}\,\Omega_{AB} \ .
\end{align}

\paragraph{Double trace operators.} 
%%%%%%%%%%%%%%%%%%%%%%%%%%%%%%%%%%%%%%%%
%
We will also need double-trace operators, which are primaries that belong to the singlet sector of the quartic tensor product of the fundamental fields. The simplest ones are labelled by two integers $n$, $s$ and are schematically 
\begin{align}
    [J_0 J_0]_{n,s}&= J_0 \square^{n} \pl^s J_0\,,
\end{align}
where  twist is $2(d-2)+2n$ and spin is $s$.  We will also encounter more complicated double-trace operators schematically of the type
\begin{align}
    [J_{s_1} J_{s_2}]_{n,s;k}&= J_{i(s_1-k)j(k)} \square^{n-k}\pl_{i(s-s_1-s_2+2k)} J_{i(s_2-k)}{}^{j(k)}\ ,
    \label{spindto}
\end{align}
whose twist is $2(d-2)+2n$, spin is $s$ and there is an additional label $k$. We use the notation $j(k) := (j_1 j_2...j_k)$. In general, we see that the spectrum of the double-trace operators is degenerate since there is more than one operator with the same twist and spin.

\paragraph{OPE and locality.}  The OPE of two $J_0$ will be a key object in what follows. It involves $2$, $1$ and $0$ Wick contractions, which schematically takes the form
\be
J_0(x_1) J_0(x_2)= 1+ \frac1{\sqrt{N}}\sum_s J_s + \sqrt{1+\frac{1}{N}} \sum_{n,s} [J_0J_0]_{n,s}
\label{OPE001}\ ,
\ee
where we have normalized the two-point function $\langle J_0(x_1) J_0(x_2) \rangle$ to be of order one. Let us denote

\begin{align}\label{ABC}
   A\equiv G_{12}G_{23}G_{34}G_{41},\quad
   B\equiv G_{13}G_{34}G_{42}G_{21},\quad
   C\equiv G_{14}G_{42}G_{23}G_{13}\ .
\end{align}
Then the dictionary between the contributions to the four point correlators (\ref{fpk1}) and the terms in (\ref{OPE001}) is:
\begin{align}
    G_{12}^2G_{34}^2&: && \text{identity operator contribution}\,, \\
    G_{13}^2G_{24}^2+G_{14}^2G_{23}^2&:&& \text{order $1$ from $[J_0J_0]_{n,s}$}\,,\\
    C&: && \text{order $1/N$ from $[J_0J_0]_{n,s}$}\,,\label{ABC2}\\
    A+B&: && \text{single-trace operators $J_s$}\,.
\end{align}
A useful byproduct of this analysis is that we can easily see the nonlocality of higher spin theories. Indeed, the s-exchange in AdS gives $A+B$ and, therefore, the sum over s-, t- and u-channels,  gives $2(A+B+C)$. The overall factor of $2$ is problematic since the correct result for the complete connected correlator should be $A+B+C$. This means that to match the bulk computation with the boundary one, one needs the quartic vertex to subtract an extra $A+B+C$ contribution. However, these contributions are of the same type as singular parts of bulk exchanges, which leads us to conclude that the quartic vertex in the higher spin theory has to be of the same degree of nonlocality as the sum of exchanges \cite{Sleight:2017pcz}. 

This is reminiscent of closed string field theory \cite{Saadi:1989tb} where, depending on the choice of cubic interaction vertices, the exchange diagram can give the right answer for the four-point function, but with a wrong prefactor. The quartic vertex is then highly non-local and is set to compensate for this over- or under-counting.  What saves the day is the worldsheet interpretation, which is missing in higher spin gravities at present. We would encounter the same problem in string field theory if tried to decompose the  diagrams into sums over fields with definite spin and mass.

%%%%%%%%%%%%%%%%%%%%%%%%%%%%%%%%%%%%%%%%%%%%%%%%%%%%%%%%%%
\section{One loop in higher spin gravity}
\label{sec4}
%%%%%%%%%%%%%%%%%%%%%%%%%%%%%%%%%%%%%%%%%%%%%%%%%%%%%%%%%%

In this section we will apply the formula (\ref{30may6}) to the simplest higher spin gravities that are dual to free $O(N)$, $U(N)$ and $USp(N)$ vector models.
Our aim is to compute the singular part of the one-loop four-point amplitude of scalar fields, which is associated with a double-particle cut. Naturally, we need to take into account contributions from an infinite tower of higher spin fields running in the loop.  To be more precise, we will compute
\begin{equation}
{\rm disc}\left[A_4^{{\rm 1-loop}}(u,v)\right] \equiv \sum_{s_1,s_2}\sum_{n,s} \frac{\left(\delta a^{00,s_1s_2}_{{n,s}}\right)^2 }{\bar a_{{n,s}}} \  G^{00,00}_{{n,s}} (u,v).
\label{30may6x1}
\end{equation}
Here $\delta a$ refers to the conformal block coefficients of the tree-level four-point function in the higher spin theory, while $\bar a$ are conformal block coefficients for the disconnected correlator. As before, $n$ and $s$ refer to contributions of operators of twist $2(d-2)+2n$. 
While we shall focus on the computation of the double-particle cut in the one-loop diagram in the rest of this section, later on we shall deal with certain single-cut diagrams that contribute to the full result for the singular part of the one-loop amplitude as well. These diagrams as well as certain open problems pertinent to the full result will be discussed in section \ref{sec:towards}.

\begin{figure}
\centering
\includegraphics{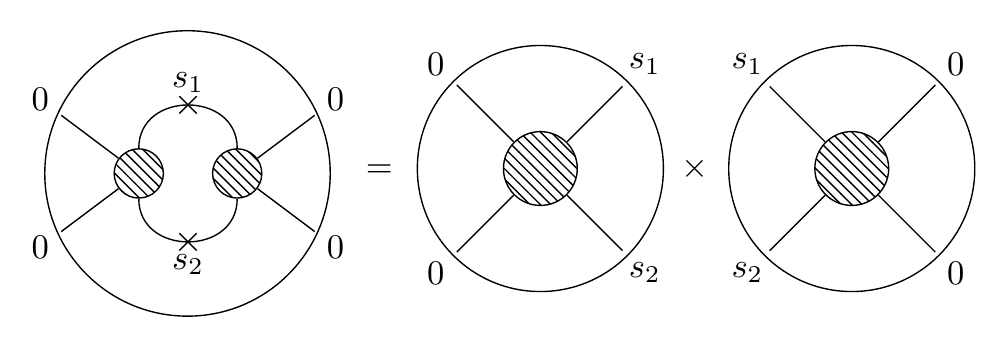}
\caption{The double cut in one-loop Witten diagram for the scattering of four scalar fields in the HSGRA is expressed in terms of tree-level on-shell amplitudes. The shaded blobs indicate the sum off all tree-level sub-diagrams. }
\label{dc}
\end{figure}

As explained in the introduction, we assume that the tree-level data for bulk higher spin theories is defined by the CFT correlators. To be able to compute the singular part of the one-loop four-point scalar amplitude with (\ref{30may6x1}), we need to find the conformal block coefficients for the tree-level amplitudes entering the cut as well as the conformal block coefficient for the disconnected correlators, that appear in the denominator of (\ref{30may6x1}). The former are defined by the $O(N^{-1})$ part of the correlators $\langle{ J}_0 { J}_0{ J}_{s_1}{ J}_{s_2} \rangle$, and the latter --- by the $O(1)$ part of $\langle{ J}_{s_1} { J}_{s_2}{ J}_{s_1}{ J}_{s_2} \rangle$. These will be evaluated simply by performing Wick contractions. 

We begin with the connected part of  the correlators $\langle{ J}_0 { J}_0{ J}_{s_1}{ J}_{s_2} \rangle$,
which can be written as
\be
 \langle{ J}_0 { J}_0{ J}_{s_1}{ J}_{s_2} \rangle \big\vert_{N^{-1}} = \sum_{n,s} \delta a^{00,s_1s_2}_{n,s}\, G^{00,s_1s_2}_{n,s}\ .
 \label{21feb1}
 \ee
Here the sum runs over all operators that appear simultaneously in the OPE of $J_0J_0$ and $J_{s_1}J_{s_2}$. Potentially, all double-trace operators of the schematic form (\ref{spindto}) can contribute to these OPE's. It is easy to see that the spectrum of these operators is degenerate, as there is finitely many of them with the same dimension and spin. Moreover,  their two-point functions, typically, are not diagonal. We shall diagonalize the sets of degenerate operators in such a way  that the diagonal basis of operators features $[J_0 J_0]_{n,s}$ as the first basis element. Then, as one can see by explicit computations, the remaining basis elements have vanishing three-point correlators with $J_{0}$ and $J_{0}$. This means that to extract the conformal block coefficients in (\ref{21feb1}) it is sufficient to compute three-point correlators $\langle J_0 J_0 [J_0J_0]_{n,s} \rangle$ and $\langle J_{s_1} J_{s_2} [J_0J_0]_{n,s} \rangle$. By conformal invariance these are proportional to some standard conformally invariant structures \cite{Mack:1976pa,Sotkov:1976xe,Osborn:1993cr,Costa:2011mg}. With the operators properly normalized, the proportionality coefficients in these correlators give the OPE coefficients $c_{J_0 J_0 [J_0,J_0]_{n,s}}$ and $c_{J_{s_1}J_{s_2}[J_0,J_0]_{n,s}}$. Then, the complete conformal block coefficient for $\langle{ J}_0 { J}_0{ J}_{s_1}{ J}_{s_2} \rangle$ is defined by 
\be
a^{00,s_1s_2}_{n,s} = c_{J_0 J_0 [J_0J_0]_{n,s}}  c_{J_{s_1}J_{s_2}[J_0J_0]_{n,s}}.
\label{21feb2}
\ee
Finally, by extracting its $O({N}^{-1})$ part, we get the conformal block coefficient $\delta a$ for the connected part of $\langle{ J}_0 { J}_0{ J}_{s_1}{J}_{s_2} \rangle$ that enters (\ref{21feb1}) 
\be
\delta a^{00,s_1s_2}_{n,s} = \left(c_{J_0 J_0 [J_0J_0]_{n,s}}  c_{J_{s_1}J_{s_2}[J_0J_0]_{n,s}}\right)\Big|_{N^{-1}}.
\label{21feb2x}
\ee

To extract the conformal block coefficients for the disconnected part of $\langle{J}_{s_1} { J}_{s_2}{J}_{s_1}{ J}_{s_2} \rangle$ we note that the only  operators that enter 
 the OPE of $J_{s_1}$ and $J_{s_2}$ with order-$O(1)$ OPE coefficients
 are of the form $[J_{s_1}J_{s_2}]_{n,s}$. Consequently, 
\begin{equation}
   \bar a_{n,s}= 
   \left(c_{J_{s_1} J_{s_2} [J_{s_1}J_{s_2}]_{n,s}}  \right)^2\Big|_{N^{0}}\ .
    \label{21feb2x1}
\end{equation}
Below, we will find it convenient to use the notation
\be
a_{J_{s_1}J_{s_2}[J_{s_1}J_{s_2}]_{n,s}} = \left(c_{J_{s_1}J_{s_2}[J_{s_1}J_{s_2}]_{n,s}}\right)^2
\,.
\ee
Summarizing,  in terms of the OPE coefficients (\ref{30may6x1}) reads
\begin{equation}
{\rm disc}\left[A_4^{{\rm 1-loop}}(u,v)\right] = \sum_{s_1,s_2}\sum_{n,s} \frac{ a_{J_0,J_0, [J_0J_0]_{n,s}}a_{J_{s_1},J_{s_2}, [J_0J_0]_{n,s}}\Big|_{N^{-2}}}{a_{J_{s_1},J_{s_2}, [J_{s_1}J_{s_2}]_{n,s}}\Big|_{N^0}} \  G^{00,00}_{{n,s}} (u,v).
\label{30may6xx1}
\end{equation}

By higher spin symmetry, we expect that the cut of the loop diagram should either vanish or be proportional to the tree-level result, i.e. to $\langle J_0J_0J_0J_0\rangle$. This four-point function has two different contributions from the $[J_0J_0]_{n,s}$ double-trace operators: the disconnected one of order $N^0$ and the connected one of order $N^{-1}$, {which we have denoted by $C$ in \eqref{ABC2}}. For these two parts we reserve the following notation 
\begin{align}
    A_{{n,s}}^0&=a_{J_0,J_0, [J_0J_0]_{n,s}}\Big|_{N^{0}}\,, &
    A_{{n,s}}^1&=a_{J_0,J_0, [J_0J_0]_{n,s}}\Big|_{N^{-1}}\,.
\end{align}
Thus, we expect the result for (\ref{30may6x1}) to take the form
\begin{align}\label{mainsystem}
    \sum_{s_1,s_2}\frac{ a_{J_0,J_0, [J_0J_0]_{n,s}}a_{J_{s_1},J_{s_2}, [J_0J_0]_{n,s}}\Big|_{N^{-2}}}{a_{J_{s_1},J_{s_2}, [J_{s_1}J_{s_2}]_{n,s}}\Big|_{N^0}}&= \alpha A_{{n,s}}^0 +\beta A_{{n,s}}^1\,.
\end{align}

Note that these are infinitely many equations labelled by $n$ and $s$ that need to be solved for  just two unknowns $\alpha$ and $\beta$. Thus, we have an infinitely over-determined system of equations. Instead of solving all of them, to specify $\alpha$ and $\beta$, it suffices to solve only two simplest equations with lowest spins and twists. Once this is done, we will then look at few other equations and verify that $\alpha$ and $\beta$ solve them as well, hence, confirming our expectation that the cut diagram is higher spin invariant. Then we will argue that the complete cut diagram is just a linear combination of the connected and disconnected parts of the boundary correlator with the proportionality coefficients given by $\alpha$ and $\beta$. It is worth to emphasize that by proceeding as explained above we do not assume higher spin symmetry. Indeed, (\ref{mainsystem}) allows to compute the right hand side for $n$ and $s$ irrespectively of the result being higher spin invariant or not. Instead, we rather find that the cut diagram we are computing is higher spin invariant and then to simplify computation focus only on its  lowest spin and lowest twist contributions and reconstruct the remaining terms by employing higher spin symmetry.

\paragraph{$\boldsymbol{O(N)}$ case.} We take the first few double-trace operators $[J_0J_0]_{n,s}$ and compute their contribution to the cut. The explicit expressions for these double-trace operators can be found in Appendix A. The very first operator $[J_0J_0]_{0,0}$ mixes with $s_1=s_2=0$ higher spin currents only, and consequently we have
\begin{align}
    \frac{ a_{J_0,J_0, [J_0J_0]_{0,0}}a_{J_0,J_0, [J_0J_0]_{0,0}}\Big|_{N^{-2}}}{a_{J_0,J_0, [J_0J_0]_{0,0}}\Big|_{N^0}}&= \alpha A_{{0,0}}^0 +\beta A_{{0,0}}^1\ .
    \label{e1}
\end{align}
The lowest twist spin-two operator $[J_0J_0]_{0,2}$ mixes with $J_0 J_0$ and $J_0 J_2$
\begin{align}
    \frac{ a_{J_0,J_0, [J_0J_0]_{0,2}}a_{J_0,J_0, [J_0J_0]_{0,2}}\Big|_{N^{-2}}}{a_{J_0,J_0, [J_0J_0]_{0,2}}\Big|_{N^0}}+
    \frac{ a_{J_0,J_0, [J_0J_0]_{0,2}}a_{J_0,J_2, [J_0J_0]_{0,2}}\Big|_{N^{-2}}}{a_{J_0,J_2, [J_0J_2]_{0,2}}\Big|_{N^0}}&= \alpha A_{{0,2}}^0 +\beta A_{{0,2}}^1\,.
\end{align}
Note that the denominator of the second term contains a different double-trace operator with the same twist and spin, $[J_0J_2]_{0,2}$. It is the only one that gives the order $N^0$ contribution. 

The next to the lowest twist operator $[J_0J_0]_{1,0}$ does not mix with anything but $J_0 J_0$ and its contribution is
\begin{align}
    \frac{ a_{J_0,J_0, [J_0J_0]_{1,0}}a_{J_0,J_0, [J_0J_0]_{1,0}}\Big|_{N^{-2}}}{a_{J_0,J_0, [J_0J_0]_{1,0}}\Big|_{N^0}}&= \alpha A_{{1,0}}^0 +\beta A_{{1,0}}^1\,.
\end{align}
Lastly, we would like to add the $n=2$ operator $[J_0J_0]_{2,0}$
\begin{align}
    \frac{ a_{J_0,J_0, [J_0J_0]_{2,0}}a_{J_0,J_0, [J_0J_0]_{2,0}}\Big|_{N^{-2}}}{a_{J_0,J_0, [J_0J_0]_{2,0}}\Big|_{N^0}}+
    \frac{ a_{J_0,J_0, [J_0J_0]_{2,0}}a_{J_2,J_2, [J_0J_0]_{2,0}}\Big|_{N^{-2}}}{a_{J_2,J_2, [J_2J_2]_{2,0}}\Big|_{N^0}}
    = \alpha A_{{2,0}}^0 +\beta A_{{2,0}}^1\,.
\end{align}
Note that $\langle J_0 J_2 [J_0J_0]_{2,0}\rangle\equiv0$ and there is no such contribution above.

The system is over-determined and may not have any solution at all. Another possible source of failure is if the tensor structures in some of the three-point functions above do not match. For example, ${\langle J_0\,J_2\, [J_0J_2]_{0,2}\rangle}$ and $\langle {J_0\,J_2\, [J_0J_0]_{0,2}}\rangle $ may have had different tensor structures, which would have made the expression meaningless. Therefore, it is a highly nontrivial statement that the system above makes sense and has a unique solution $\alpha=2$, $\beta=1$.

\paragraph{$\boldsymbol{U(N)}$ case.} The $U(N)$ case is similar to the $O(N)$ one, but there are more terms and mixings between the operators in general since we also have odd spin currents. The very first equation \eqref{e1} is unchanged as compared to the $O(N)$ case (the OPE coefficients do change, but not the structure)
\begin{align}
    \frac{ a_{J_0,J_0, [J_0J_0]_{0,0}}a_{J_0,J_0, [J_0J_0]_{0,0}}\Big|_{N^{-2}}}{a_{J_0,J_0, [J_0J_0]_{0,0}}\Big|_{N^0}}&= \alpha A_{{0,0}}^0 +\beta A_{{0,0}}^1\,.
\end{align}
The lowest twist spin-two double trace operator $[J_0J_0]_{0,2}$ receives more contributions as compared to the $O(N)$ case
\begin{align}
    \frac{ a_{J_0,J_0, [J_0J_0]_{0,2}}a_{J_0,J_0, [J_0J_0]_{0,2}}\Big|_{N^{-2}}}{a_{J_0,J_0, [J_0J_0]_{0,2}}\Big|_{N^0}}+
    \frac{ a_{J_0,J_0, [J_0J_0]_{0,2}}a_{J_0,J_2, [J_0J_0]_{0,2}}\Big|_{N^{-2}}}{a_{J_0,J_2, [J_0J_2]_{0,2}}\Big|_{N^0}}+\\
    +
    \frac{ a_{J_0,J_0, [J_0J_0]_{0,2}}a_{J_1,J_1, [J_0J_0]_{0,2}}\Big|_{N^{-2}}}{a_{J_1,J_1, [J_1J_1]_{0,2}}\Big|_{N^0}}= \alpha A_{{0,2}}^0 +\beta A_{{0,2}}^1\,.
\end{align}
The next to the lowest twist operator $[J_0J_0]_{1,0}$ receives an additional contribution due to the spin-one current $J_1$
\begin{align}
    \frac{ a_{J_0,J_0, [J_0J_0]_{1,0}}a_{J_0,J_0, [J_0J_0]_{1,0}}\Big|_{N^{-2}}}{a_{J_0,J_0, [J_0J_0]_{1,0}}\Big|_{N^0}}+
    \frac{ a_{J_0,J_0, [J_0J_0]_{1,0}}a_{J_1,J_1, [J_0J_0]_{1,0}}\Big|_{N^{-2}}}{a_{J_1,J_1, [J_1J_1]_{1,0}}\Big|_{N^0}}&= \alpha A_{{1,0}}^0 +\beta A_{{1,0}}^1\,.
\end{align}
Note that $\langle J_0 J_1 [J_0J_0]_{1,0}\rangle\equiv0$ and there is no such contribution above. Lastly, we consider the $n=2$  operator 
\begin{align}
    \frac{ a_{J_0,J_0, [J_0J_0]_{2,0}}a_{J_0,J_0, [J_0J_0]_{2,0}}\Big|_{N^{-2}}}{a_{J_0,J_0, [J_0J_0]_{2,0}}\Big|_{N^0}}+
    \frac{ a_{J_0,J_0, [J_0J_0]_{2,0}}a_{J_2,J_2, [J_0J_0]_{2,0}}\Big|_{N^{-2}}}{a_{J_2,J_2, [J_2J_2]_{2,0}}\Big|_{N^0}}+\\
    +
    \frac{ a_{J_0,J_0, [J_0J_0]_{2,0}}a_{J_1,J_1, [J_0J_0]_{2,0}}\Big|_{N^{-2}}}{a_{J_1,J_1, [J_1J_1]_{2,0}}\Big|_{N^0}}
    = \alpha A_{{2,0}}^0 +\beta A_{{2,0}}^1\,.
\end{align}
Again the system is over-determined, but the equations are satisfied for $\alpha=1$ and $\beta=0$.

\paragraph{$\boldsymbol{USp(N)}$ case.} The pattern above --- over-determined system of equations which admits a solution --- seems to be convincing and we consider two equations only in order to fix $\alpha$ and $\beta$ for the $USp(N)$-case. The first equation is
\begin{align}
    \frac{ a_{J_0,J_0, [J_0J_0]_{0,0}}a_{J_0,J_0, [J_0J_0]_{0,0}}\Big|_{N^{-2}}}{a_{J_0,J_0, [J_0J_0]_{0,0}}\Big|_{N^0}}&= \alpha A_{{0,0}}^0 +\beta A_{{0,0}}^1\,.
\end{align}
The second equation corresponds to $[J_0J_0]_{1,0}$
\begin{align*}
    \frac{ a_{J_0,J_0, [J_0J_0]_{1,0}}a_{J_0,J_0, [J_0J_0]_{1,0}}\Big|_{N^{-2}}}{a_{J_0,J_0, [J_0J_0]_{1,0}}\Big|_{N^0}}+
    \sum_{i\leq j}\frac{ a_{J_0,J_0, [J_0J_0]_{1,0}}a_{J_1^i,J^j_1, [J_0J_0]_{1,0}}\Big|_{N^{-2}}}{a_{J_1^i,J^j_1, [J_1J_1]^{ij}_{1,0}}\Big|_{N^0}}&= \alpha A_{{1,0}}^0 +\beta A_{{1,0}}^1\,,
\end{align*}
where we take into account contributions from the triplet of spin-one currents $J^{ab}_1$.
Their two-point function $\langle J^{ab}_\mu J^{a'b'}_\nu\rangle$  has an additional factor $\epsilon^{aa'}\epsilon^{bb'}+\epsilon^{ab'}\epsilon^{a'b}$. We diagonalize it by choosing $J^i$, $i=1,2,3$ appropriately. Different contributions are then summed over. Also note that only order $N^0$ term is needed in the denominator. It comes from a specific double-trace operator, $[J_1J_1]^{ij}_{1,0}$, constructed out $J^i$ and $J^j$. The system has a unique solution $\alpha=2$ and $\beta=-1$.

\paragraph{Applicability of the cutting formula.}
The method we have used to compute double-particle singularities is motivated by a direct extension of the formula (\ref{30may6}) which was proven in the context of a toy model that does not capture a number of features in higher spin gravity. In this model, the double-trace operators  associated with  external lines, $[\phi\phi]$ and $[\psi\psi]$, have different quantum numbers compared to the double-trace operators associated with fields running in the loop, $[\chi\chi]$. In this setup it was shown that the cut contributions are captured by $[\chi\chi]$ double-trace conformal blocks and the conformal block coefficients are given by the formula (\ref{30may6}). In contrast, in our case double-trace operators associated with external lines, $[J_0J_0]_{n,s}$, and those associated with fields running in the loop, $[J_{s_1}J_{s_2}]_{n,s}$, contain operators of the same dimension and spin. Normally, under these circumstances one expects presence of anomalous dimensions and, as a result, derivatives of the conformal blocks in the conformal block decomposition of both tree-level and loop-level diagrams, see e.g. \cite{Aharony:2016dwx}. However, due to duality with free CFT's, in the case of HSGRA such anomalous dimensions are not present at tree-level and are not expected at loop-level, which we confirm by our analysis. Therefore, application of the method of \cite{Fitzpatrick:2011dm} even to the higher spin theory case seems reasonable.

Another potential issue with the direct application of the cutting formula (\ref{30may6}) to the higher spin case is that it does not take into account the contributions of single-trace conformal blocks to tree-level diagrams. In the toy model, where this formula was derived, such contributions were absent due to absence of cubic interactions in the bulk. However, they do arise in the higher spin theory and may, in principle, contribute to the cut diagram we have computed. To the best of our knowledge, a generalization of the formula (\ref{30may6}) that would take into account tree-level single-trace contributions to the double-cut diagram is not available so far. While we expect that the single-trace part of tree-level diagrams do not contribute to the double-cut loop diagram we have computed, this issue remains to be investigated\footnote{{This expectation is confirmed by the analysis of \cite{Ponomarev:2019ofr}. Similarly, in \cite{Ponomarev:2019ofr} it was shown that (\ref{30may6}) can be used to compute the double-particle cut diagram in the higher-spin case despite the absence of anomalous dimensions.} }.

%%%%%%%%%%%%%%%%%%%%%%%%%%%%%%%%%%%%%%%%%%%%%%%%%%%%%%%%%%%%%
\section{Towards the complete four-point one-loop amplitude}
\label{sec:towards}
%%%%%%%%%%%%%%%%%%%%%%%%%%%%%%%%%%%%%%%%%%%%%%%%%%%%%%%%%%%%%

In the previous section we computed the s-channel double-particle cut contribution to the one-loop four-point amplitude in higher spin theories dual to $U(N)$, $O(N)$ and $USp(N)$ free vector models on the boundary. The result can be summarized as follows 
\begin{align}
\label{22feb1}
\sum_{s_1,s_2}{\parbox{3.6cm}{\includegraphics{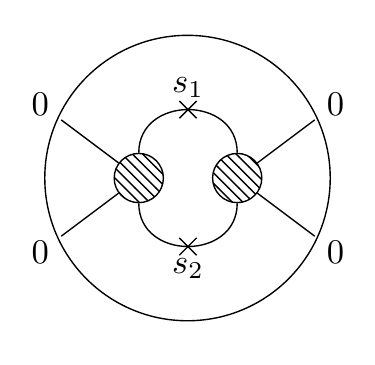}}} = \alpha \,\text{Disconnected} +\beta\, \text{Connected}
\end{align}
where 
\be
(\alpha,\beta) = \begin{cases} (2,+1) & \mbox{for} \quad O(N)\ ,\\ (1,0) & \mbox{for} \quad U(N) \ ,\\ (2,-1) & \mbox{for} \quad USp(N)\ .\end{cases}
\label{ab}
\ee
Here by {\it Connected} and {\it Disconnected} we mean the contribution of the double-trace operators to the connected and disconnected components of the four-point function $\langle J_0J_0J_0J_0\rangle$.
{Note that unlike in more typical bulk theories, the double-cut diagram (5.1) does not lead to any anomalous dimensions, which is a consequence of the duality with the free CFT}
As we explained in the introduction, our main motivation in carrying out this analysis is to test higher spin holography for four-point functions at one-loop level. To be able to do that we should reconstruct the complete bulk one-loop amplitude. Before discussing which terms are not yet captured by (\ref{22feb1}) and how they can be accessed we first review the expected result for this computation, which follows from the analysis of vacuum one-loop diagrams. 

\paragraph{Expectations from one-loop vacuum diagrams.}

Let us assume that we have a complete and background independent action for the higher spin gravity with $\Phi$ being the collective notation for the whole multiplet
\begin{align}
    S&= \frac{1}{G} \int d^dx\, \mathcal{L}(\Phi)\ .
\label{action3}
\end{align}
We take the vacuum solution $\Phi_0$ to be $AdS_d$ and expand $\Phi\rightarrow\Phi_0+G^{1/2}\Phi$:
\begin{align}
    S&= \frac{1}{G}S_0+ \int \tfrac12  \Phi K\Phi +G^{1/2} V_3(\Phi,\Phi,\Phi)+G V_4(\Phi,\Phi,\Phi,\Phi)+ 
\end{align}
Then the propagator scales as $G^0$ and the $n$-point vertex scales as $G^{(n-2)/2}$. As a consequence, at tree level the connected contribution to the four-point holographic correlation function scales as $G$ and the disconnected one scales as $G^0$. Next, the large-$N$ approximation suggests that $G^{-1}\sim N$. Moreover, $N$ is expected to be quantized \cite{Maldacena:2011jn}, which is hard to see from the bulk. This explains our choice of normalization for the CFT correlation functions. 

Ideally, one should compute the complete path integral on both sides of the duality. In particular, on the $AdS$-side we have schematically
\begin{align}
    Z_{AdS}&= \int D\Phi\, e^{-S[\Phi]}\,.
\end{align}
The free energy (as well as the full partition function with the sources turned on) admits an expansion in the coupling constant $G$
\begin{align}
    -\ln Z_{AdS}&=F_{AdS}=\frac{1}{G} F^0_{AdS} +F^1_{AdS} +G F^2_{AdS}+...\,,
\end{align}
where the first term is the classical action evaluated on the (Euclidean) $AdS_{d+1}$ background, the term we do not have access to; $F^1$ corresponds to one-loop corrections, etc.  On the dual CFT side there should be a similar expansion for the CFT free energy $F_{CFT}$:
\begin{align}
    -\ln Z_{CFT}&=F_{CFT}=N F^0_{CFT}+F^{1}_{CFT}+\frac1{N}F^2_{CFT}+...\,.
\end{align}
For the free CFT dual $F^{1}_{CFT}$, as well as all higher orders, vanish. This does not have to be so in the bulk. Indeed, the relation $G(N)$ between $G$ and $N$ may be more complicated than just $G^{-1}\sim N$.
In fact, one finds that the one-loop correction for the dual of the free $U(N)$ vector model does vanish, consistently with the simplest relation $G^{-1}\sim N$. Instead, for the dual of the free $O(N)$ vector model $F^1_{AdS}=F^0_{CFT}$ \cite{Giombi:2013fka}. Therefore, the duality suggests that $G^{-1}\sim N-1$ at one-loop level and the missing $1\times F^0_{CFT}$ results from the one-loop correction $F^1_{AdS}$. Similarly, for the $USp(N)$-model $F^1=- F^0_{CFT}$, which suggests $G^{-1}\sim N+1$. These results hold true in all dimensions\footnote{We note, however, that there is a puzzle with the one-loop corrections for the Type-B higher spin theory in all $AdS_{2n}$, \cite{Giombi:2013fka,Gunaydin:2016amv,Giombi:2016pvg}.} $d$ and can even be extended to non-integer ones \cite{Skvortsov:2017ldz} in accordance with the fact that the vector models can be defined in fractional dimension within the large-$N$ expansion scheme. Moreover, it was suggested \cite{Klebanov:2002ja} that the duality may extend to $AdS_{5-\epsilon}/CFT_{4-\epsilon}$, which is especially interesting for the critical vector model. To sum up, depending on the case of the AdS/CFT duality we consider, the vacuum corrections lead to the relation 
\be\label{GN}
G^{-1}\sim N+b \quad {\rm with}\quad b=\begin{cases} -1& {\rm for} \quad O(N)\,, \\ 0 & \mbox{for} \quad U(N)\,, \\ 1 & \mbox{for}\quad  USp(N)\,. \end{cases}
\ee

If holography is to hold, once the shift in the identification between the bulk coupling constant and $1/N$ found for vacuum diagrams is taken into account, higher-point bulk amplitudes computed to one-loop order should agree with the respective free correlators on the boundary.  A simple computation shows that this implies that the connected one-loop four-point bulk  amplitude should be equal to 
\begin{equation}
    \label{28mar1}
    b\cdot  G^2 (A+B+C),
\end{equation}
where $A$, $B$ and $C$ were defined in (\ref{ABC}). Moreover, disconnected bulk diagrams should agree with disconnected CFT correlators as a result of matching of two-point functions.

\paragraph{Towards the complete amplitude.} We will now proceed with the discussion of how the complete one-loop amplitude in the higher spin theory can be reproduced once the double-cut diagram (\ref{22feb1}) is known.  When computing the amplitude using the on-shell methods, the strategy is to compute its singular part and then reproduce a complete amplitude by requiring the appropriate high energy behaviour, presence of all necessary singularities and crossing symmetry. Due to a peculiar nature of higher spin theories we encounter a number of unusual features, which prevent us from applying this strategy directly. Moreover, there are some genuine limitation of the on-shell approach that do not allow us reconstruct the amplitude completely. Let us discuss these issues. 

The most obvious unexpected feature of the result for the double-cut diagram (\ref{22feb1}) is that it contains a disconnected contribution.  While the very fact that summation over infinitely many spins may turn connected diagrams into disconnected ones is unusual, though, not totally surprising,\footnote{In fact, other unconventional effects have been observed in higher spin gravities: in the conformal HSGRA \cite{Joung:2015eny,Beccaria:2016syk} the sum over exchange diagrams produces a delta-function of the Mandelstam variables which is not present in each of the diagrams. Similarly, the Mellin amplitude of the holographic HSGRA, from bulk perspective, takes a rather unusual form -- depending on how various subtleties with its definition in a given case are resolved it is either distributional or vanishes \cite{Taronna:2016ats,Bekaert:2016ezc,Rastelli:2017udc,Ponomarev:2017qab}.} this effect has more significant consequences, as it 
can destroy the holographic duality, because the four-point function would be inconsistent with the OPE. Therefore, we have to assume that such disconnected contributions will eventually be cancelled, if higher spin/AdS holography is to hold. 

Next, the double-particle cut we computed does not capture the complete $s$-channel singularity of the one-loop amplitude: single-cut singularities should also be taken into account. At a given order, a single-particle singularity factorizes the amplitude into a product of tree-level and one-loop three-point amplitudes.  The latter cannot be computed by the on-shell methods\footnote{{The reason is that three-point amplitudes are not given by functions of some independent kinematic variables, but, rather, by a single real number (or a set of real numbers, for the case of spinning fields, for which several tensor structures are available).  So it does not make sense to talk about their analytic properties and, hence, techniques based on these analytic properties are inapplicable. }} and should be taken as an external input into the procedure. We may, however, assume that one-loop three-point amplitudes (the blob denotes the one-loop vertex correction)
\begin{align}
    W_3&= G^{1/2}\parbox{2.0cm}{\includegraphics{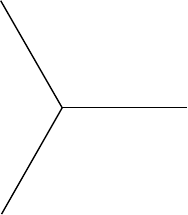}} + G^{3/2} \parbox{2.0cm}{\includegraphics{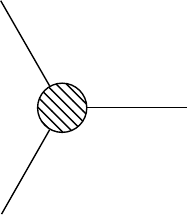}}+... \, .
\end{align}
 are as required by matching with the boundary three-point correlators. This implies that the one-loop three-point amplitude should be equal to $b/2$ of the tree-level result.\footnote{We should have $N^{-1/2}=G^{1/2}+\gamma G^{3/2}$ where $\gamma$ is the one-loop correction. This implies, $\gamma=b/2$ at the first nontrivial order.} This, eventually, entails  that the $s$-channel single-particle cut diagram is given by

{\begin{equation}
    \label{28mar2}
    2 \gamma G^{3/2} G^{1/2} (A+B) = b\cdot G^2 (A+B)\,.
\end{equation}}
{To achieve this result we use the tree-level computation where one of the two vertices is corrected by the quantum factor of $\gamma G$, moreover, each of the two vertices can be corrected --- hence, the factor of 2.}

Adding to it a double-particle singularity, we find that the complete $s$-channel singularity is given by ($A$, $B$ and $C$ are defined in (\ref{ABC}))
\begin{equation}
    \label{28mar3}
    \beta\cdot G^2 C +  b\cdot G^2 (A+B) = b\cdot G^2 (A+B-C),
\end{equation}
where we used that in all cases that we considered, we found $\beta=-b$.

As a next step, we are supposed to find a crossing symmetric expression that contains  $s$-channel singularity (\ref{28mar3}). Unlike in more conventional field theories, where different cut diagrams result into different singularities, we find that in the higher spin case singularities mix up. For example, the contribution $A+B$ can be either interpreted as a single-particle singularity in the $s$-channel or as a sum of two double-particle singularities in crossed channels. To resolve this mixing problem one needs a better understanding of how the methods based on analyticity and unitarity extend to the case of non-local theories, such as higher spin theory.\footnote{For discussion of Cutkosky rules in string field theory see \cite{Pius:2016jsl}.}  We just note here that if we naively symmetrize the $s$-channel singularity (\ref{28mar3}) over channels

{\begin{align}
    b G^2[(A+B-C) + (B+C-A)+(C+A-B)] =b G^2 (A+B+C) 
\end{align}}
we find the desired result (\ref{28mar1}) and the so obtained complete one-loop correction is consistent with the duality --- it is consistent with $G(N)$ \eqref{GN} that has been already fixed by the vacuum one-loop corrections.

Once the singular part is known one can reconstruct the complete amplitude using AdS dispersion relations, see \cite{Alday:2016njk,Caron-Huot:2017vep,Alday:2017vkk}. For the case at question, however, the singular contribution already gives a correlator of a consistent CFT, which satisfies all necessary conditions and, hence, does not need any completion. Alternatively, one can notice that when dimensions of fields on external lines are equal, regular terms produced by dispersion relations give rise to contributions involving derivatives of conformal blocks. These, however, are inconsistent with the exact higher spin symmetry, because the latter implies absence of anomalous dimensions and, in fact, fixes all correlation functions  \cite{Maldacena:2011jn,Boulanger:2013zza,Alba:2013yda,Alba:2015upa}. A similar phenomenon also occurs at tree level, where the exchange singularities contribute to the complete amplitude without any dressing with regular terms. In other words, we expect that,  contrary to more standard field theories, in the higher spin case singularities already give a complete amplitude.

Another potential source of contributions, that have not yet been taken into account, originates from self-energy corrections:
\begin{align}
    \langle J J\rangle&=W_2\,, & W_2&=  G^{0}\includegraphics{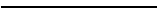} +
   G^1\raisebox{-0.5cm}{\includegraphics{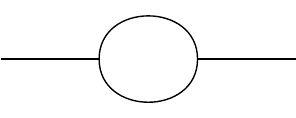}}+...\,.%\diagProptwoloops
\end{align}
 To be more precise, similarly to the flat-space $S$-matrix, the definition of the holographic $S$-matrix involves the appropriate LSZ reduction.\footnote{See \cite{Giddings:1999qu} for a step in that direction. See also \cite{Dusedau:1985ue} for a discussion of the Kallen-Lehmann representation in AdS.} This procedure requires the knowledge of the bulk-to-bulk propagator to one-loop order. Though, it is not available,\footnote{In AdS gravity coupled to $\phi^4$ the self-energy has been studied in detail in \cite{Bertan:2018khc,Bertan:2018afl} up to two loops.} higher spin symmetry implies that the only effect that loop corrections produce is that of the wave-function renormalization, $G_2\to Z G_2$. Then, to match the bulk two-point function with the boundary  correlator, which has a unit normalization, we will have to rescale the bulk fields. The account of this contribution requires the knowledge of $Z$ and we leave it for future research.

%%%%%%%%%%%%%%%%%%%%%%%%%%%%%%%%%%%%%%%%%%%%%%%%%%%%%%%%%%%%%
\section{Conclusions and Future Directions}
\label{sec:Discussion}
%%%%%%%%%%%%%%%%%%%%%%%%%%%%%%%%%%%%%%%%%%%%%%%%%%%%%%%%%%%%%
The main 
%solid 
result of the paper is the computation of the double-cut of the one-loop correction to the four-point function in the higher spin gravities dual to $U(N)$, $O(N)$ and $USp(N)$ free vector models. With certain assumptions regarding the other contributions, in particular, those coming from the self-energy corrections and from the vertex corrections, we find evidence for matching of the one-loop correction to the four-point function with the CFT result, upon performing the shift in the inverse gravitation coupling constant $G^{-1}$ derived from the one-loop correction to the vacuum energy. While this is a highly nontrivial test, it should be taken with a grain of salt since there is no complete control over all one-loop corrections. The most puzzling effect is the appearance of the disconnected contributions that have to be cancelled eventually.

Let us stress that the approach pursued in the present paper is to reduce certain bulk loop computations to well-defined manipulations with the CFT data, rather than that of merely interpreting higher order corrections to the CFT correlation functions as certain computations in the dual gravitational theory. Nevertheless, there are important effects that are not probed by our tool, e.g. it is important to see that all UV divergences and sums over infinitely many fields running in the loops lead to finite answers after an appropriate regularization, like they do for the vacuum one-loop corrections.

There is a number of obvious directions that the present results can be extended to. We have probed the simplest corrections to the simplest types of the higher spin AdS/CFT duality that are based on the free boson CFT's. 
Optimistically, better understanding of the cutting rules and  procedures enabling reconstruction of complete amplitudes form singularities  in the higher spin case should allow one to extend the present results to all orders in $G$ and determine the exact dependence of the bulk coupling constant $G$ on $N$. There are many other free CFT's that the present results can be extended to. In particular, it is interesting to extend it to the free fermion CFT and to see whether it gives a clue to the discrepancy observed for vacuum corrections in \cite{Giombi:2013fka,Gunaydin:2016amv,Giombi:2016pvg}. Extensions to more stringy-like dualities that are based on free CFT's, e.g. free $\mathcal{N}=4$ SYM, with matter in matrix representations \cite{Sundborg:2000wp,Sezgin:2002rt,Beisert:2004di,Bae:2016rgm,Bae:2016hfy} can shed more light on the tensionless limit at the quantum level.

%%%%%%%%%%%%%%%%%%%%%%%%%%%%%%%%%%%%%%%%%%%%%%%%%%%%%%%%%%%%%
\section*{Acknowledgments}
\label{sec:Aknowledgements}
%%%%%%%%%%%%%%%%%%%%%%%%%%%%%%%%%%%%%%%%%%%%%%%%%%%%%%%%%%%%%
We are grateful to  Euihun Joung  and Ivo Sachs for  helpful discussions.  The work of E.Sk. was supported by the Russian Science Foundation grant 18-72-10123 in association with the Lebedev Physical Institute. The work of E. Se. and D. P. was supported by NSF grants PHY-1521099 and PHY-1803875. We thank the Erwin Schr{\"o}dinger Institute in Vienna for hospitality during the program 'Higher Spins and Holography' where this work was finished. E.Sk. also thanks Mitchell Institute at Texas A\&M University for its support and hospitality while this work was in progress.

%%%%%%%%%%%%%%%%%%%%%%%%%%%%%%%%%%%%%%%%%%%%%%%%%%%%%%%%%%
\begin{appendix}
\renewcommand{\thesection}{\Alph{section}}
\renewcommand{\theequation}{\Alph{section}.\arabic{equation}}
\setcounter{equation}{0}\setcounter{section}{0}
%%%%%%%%%%%%%%%%%%%%%%%%%%%%%%%%%%%%%%%%%%%%%%%%%%%%%%%%%%

%%%%%%%%%%%%%%%%%%%%%%%%%%%%%%%%%%%%%%%%%%%%%%%%%%%%%%%%%%
\section{Operators}
\label{app:FirstOrder}
\setcounter{equation}{0}
%%%%%%%%%%%%%%%%%%%%%%%%%%%%%%%%%%%%%%%%%%%%%%%%%%%%%%%%%%

Below we collect explicit formulas for the operators that are involved in the computation. 

\paragraph{Higher Spin Currents. } Single-trace operators \cite{Craigie:1983fb} are easy to built with the help of generating functions. For $O(N), U(N)$ and $USp(N)$ cases we have
\bea
O(N): && \qquad J(x,\xi) =F(\xi\cdot\pl_1,\xi\cdot \pl_2)\, :\phi^a(x_1)\phi^b (x_2): \delta_{ab} \Big|_{x_1=x_2=x}\ ,
\nn\\
U(N): && \qquad 
J(x,\xi) =F(\xi\cdot\pl_1,\xi\cdot \pl_2)\, :\bar{\phi}^a(x_1)\phi_a (x_2):  \Big|_{x_1=x_2=x}\,
\nn\\
USp(N): && \qquad J^{ab}(x,\xi) =F(\xi\cdot\pl_1,\xi\cdot \pl_2)\, :\bar{\phi}^{aA}(x_1)\phi^{bB}(x_2) \Omega_{AB}:  \Big|_{x_1=x_2=x}\ ,
\label{HScurrentscalarU}
\eea
where $J^{[ab]}$ contains all even spins and $J^{(ab)}$ contains all odd spins. Here auxiliary vector $\xi$ is light-like, $\xi \cdot \xi=0$. The function $F$ reduces to the Gegenbauer polynomials,  $F_s(w)=C^{(d-3)/2}_s(w)$. 

\paragraph{Double trace operators $O_{\Delta_1} \Box^n \pl^s O_{\Delta_2}$. } The simplest double-trace operators have the form
\begin{align}\label{JJos}
    [O_{\Delta_1}O_{\Delta_2}]_{0,l}(x,\xi)&=F(\xi\cdot\pl_1,\xi\cdot \pl_2)\, :O_{\Delta_1}(x_1)O_{\Delta_2} (x_2): \Big|_{x_1=x_2=x}\,,
\end{align}
where the function is expressed in terms of the Jacobi polynomials:
\begin{align}
    F=(\xi\cdot p_2 +\xi\cdot p_1)^s P^{\Delta_1-1,\Delta_2-1}_s(w)\,, && w=\frac{\xi\cdot p_2 -\xi\cdot p_1}{\xi\cdot p_2 +\xi\cdot p_1}\,.
\end{align}
Double-trace operators of type $[J_0J_0]_{n,s}$ were constructed in \cite{Bekaert:2015tva}. For our computations we need $[J_0,J_0]_{2,0}$, which is given by \eqref{JJos} and
\begin{align}
    [J_0,J_0]_{1,0} &= -2\square J_0 J_0 +\pl_i J_0 \pl^iJ_0\ ,\\
    {[J_0,J_0]_{2,0}}&=\frac{\Delta _0+1}{\Delta _0+2}2\square^2 J_0 J_0-\frac{4 \left(\Delta _0+1\right)}{\Delta _0}\square \pl_\mu J_0 \pl^\mu J_0 +\notag\\
    &\qquad \qquad +\frac{2 \Delta _0^2+7 \Delta _0+4}{\Delta _0^2}\square J_0\square J_0+\pl_\mu\pl_\nu J_0 \pl^\nu\pl^\mu J_0\, .
\end{align}
Expression for the most general double-trace operators are not available in the literature, but it is not hard to work out the ones we need for our computations. For example, we have ($\Delta_0=d-2$)
\bea
    [J_0J_2]_{0,2} &=& J_0 J_2^{\mu\nu}\ ,\qquad  [J_1J_1]_{1,0} =J_1^\mu J_{1\mu}\ , \qquad 
    [J_2J_2]_{2,0} =J_2^{\mu\nu} J_{2\mu\nu}\ , 
\w2
[J_1J_1]_{0,2} &=& J_1^\mu J_1^\nu-\frac{1}{d}\eta^{\mu\nu}J_1^\rho J_1^\sigma \eta_{\rho\sigma}\ ,
\w2
[J_1J_1]_{2,0} &=& \frac{2 \Delta _0 \left(\Delta _0+2\right)}{3 \Delta _0+4}\square J_\mu J^\mu-\frac{\Delta _0^2+3 \Delta _0+4}{3 \Delta _0+4} \pl_\mu J_\nu \pl^\mu J^\nu+\pl_\mu J_\nu \pl^\nu J^\mu\ .
\eea
%

%%%%%%%%%%%%%%%%%%%%%%%%%%%%%%%%%%%%%%%%%%%%%%%%%%%%%%%%%%
\section{Correlators}
\label{app:Correlators}
\setcounter{equation}{0}
%%%%%%%%%%%%%%%%%%%%%%%%%%%%%%%%%%%%%%%%%%%%%%%%%%%%%%%%%%
Below we collect the correlators that are used in the main text. In section \ref{sec:freeCFT} we specified the normalization of two-point functions. With this normalization we get for $J_s$ two-point function in the $U(N)$ case ($\Delta_0=d-2$)\footnote{We defined the standard conformal structures \cite{Costa:2011mg}: 
\begin{align*}
    P_{12}&= \frac{1}{x_{12}^2}\left (\xi_1\cdot \xi_2 -2\frac{(x\cdot \xi_1) (x\cdot \xi_2)}{x^2} \right )\,,\qquad \text{etc.} &
    Q_3&=-\frac{x_{12}\cdot \xi_1}{x_{12}^2}+\frac{x_{13}\cdot \xi_1}{x_{13}^2}\,,\qquad \text{etc.}
\end{align*}
}
\begin{align}
    \langle J_s J_s\rangle &= N\frac{1}{(x^2_{12})^{d-2}} \left(P_{12}\right)^s \times \frac{(-)^s2^{ s} \Gamma (2 s+\Delta_0-1) \Gamma (s+\Delta_0-1)}{\Gamma (\Delta_0-1)^2s! }\,.
\end{align}
In the $O(N)$ case the odd spin currents are absent  and there is an additional factor of $2$. For the $USp(N)$-case, $J_s$ acquire the $SU(2)$ indices, and the factor $(\epsilon^{a_1a_2}\epsilon^{b_1b_2}+\epsilon^{a_1b_2}\epsilon^{b_1a_2})$ on the right hand side. For example,
\be
\langle J_1^{a_1b_1} J_1^{a_2 b_2}\rangle&= 4N\Delta_0(\Delta_0-1)^2\frac{1}{(x_{12}^2)^{\Delta_0}} P_{12} (\epsilon^{a_1a_2}\epsilon^{b_1b_2}+\epsilon^{a_1b_2}\epsilon^{b_1a_2}) \ .
\ee

\paragraph{$\boldsymbol{O(N)}$-case.}
%%%%%%%%%%%%%%%%%%%%%%%%%%%%%%%%%%%%%%%%%
The OPE coefficient of $[J_0J_0]_{n,s}$ operators for the $O(N)$-model were computed in \cite{Bekaert:2015tva} and we simply state the result
\begin{align}
c^2_{n, s} &= \frac{\left[\left(-1\right)^s+1\right] 2^{s}\left(\tfrac{d}{2}-1\right)^{2}_n \left(d-2\right)^2_{s+n}}{s! n! \left(s+\tfrac{d}{2}\right)_n \left(d-3 + n \right)_n 
\left(2d + 2n +s-5\right)_{s} \left(\tfrac{3d}{2}-4 + n +s\right)_n} 
\nonumber \\ 
& \hspace*{4cm} \times \left(1+\left(-1\right)^n\frac{4}{N} \frac{\Gamma\left(s\right)}{2^{s}\Gamma\left(\frac{s}{2}\right)} \frac{\left(\frac{d}{2}-1\right)_{n+\tfrac{s}{2}}}{\left(\frac{d-1}{2}\right)_{\tfrac{s}{2}}
\left(d-2\right)_{n+\tfrac{s}{2}}}\right).
\label{c1}
\end{align}
Also all correlators of type $\langle J_0 J_0 [J_0 J_0]_{n,s}\rangle$, $\langle [J_0 J_0]_{n,s} [J_0 J_0]_{n,s}\rangle$ can be found in \cite{Bekaert:2015tva} and the lowest twist conformal block coefficients for $\langle J_0 J_0 J_{s_1}J_{s_2} \rangle$ were computed in \cite{Sleight:2018epi}. The relevant correlators are collected below 

{\allowdisplaybreaks\footnotesize\begin{align*}
    \langle J_0 J_2 [J_0J_0]_{0,2}\rangle&=16 N(\Delta_0-1) \Delta_0^2 (\Delta_0+1) (\Delta_0+2) (2 \Delta_0+1) \frac{ P_{23}^2}{(x_{13}^2)^{\Delta_0}(x_{23}^2)^{\Delta_0}}\ , 
    \nn\w2
    \langle J_0 J_2 [J_0J_0]_{0,2}\rangle&=16N(\Delta_0^2-1) \Delta_0^2 (\Delta_0+2) (1+2\Delta_0) \frac{P_{23}^2}{(x_{13}^2)^{\Delta_0}(x_{23}^2)^{\Delta_0}} \ ,
    \nn\w2
    \langle J_0 J_2 [J_0J_2]_{0,2}\rangle&=8N(\Delta_0-1)^2 \Delta_0^2 (\Delta_0+2) (\Delta_0+(\Delta_0+1) N) \frac{P_{23}^2}{(x_{13}^2)^{\Delta_0}(x_{23}^2)^{\Delta_0}} \ ,
 \nn\w2
     \langle J_2 J_2 [J_0J_0]_{2,0}\rangle&=64N (\Delta_0-1)^2\Delta_0^2(1+\Delta_0)(2+\Delta_0)(4+3\Delta_0)\frac{(P_{12}+2 Q_1 Q_2)^2 }{(x_{12}^2)^2(x_{13}^2)^{2+\Delta_0}(x_{23}^2)^{2+\Delta_0}} \ ,
     \nn\w2
    \langle J_2 J_2 [J_2J_2]_{2,0}\rangle&=16N (\Delta_0-1)^4\Delta_0^4(1+\Delta_0)(2+\Delta_0)^2(2N+\Delta_0(1+2N))
     \frac{(P_{12}+2 Q_1 Q_2)^2}{(x_{12}^2)^2(x_{13}^2)^{2+\Delta_0}(x_{23}^2)^{2+\Delta_0}} \ ,
\end{align*}
\begin{align*}
    \langle [J_0J_0]_{0,0}[J_0J_0]_{0,0}\rangle&= 8N (2+N) \frac{1}{(x_{12}^2)^{2\Delta_0}}\ , \\
     \langle [J_0J_0]_{1,0}[J_0J_0]_{1,0}\rangle&= \Delta _0^2 \left(\Delta _0+2\right) (N-1) N \frac{1}{(x_{12}^2)^{2(1+\Delta_0)}}\ , \\
     \langle [J_0J_0]_{2,0}[J_0J_0]_{2,0}\rangle&=128 \left(\Delta _0+1\right)^2 \left(\Delta _0+4\right) \left(3 \Delta _0+2\right) \left(3 \Delta _0+4\right) \left(\Delta _0+2 \left(\Delta _0+1\right) N+2\right) \frac{1}{(x_{12}^2)^{2(2+\Delta_0)}}\ , \\
     \langle [J_0J_0]_{0,2}[J_0J_0]_{0,2}\rangle&=32 N\Delta _0^2 \left(\Delta _0+1\right){}^2 \left(2 \Delta _0+1\right) \left(N(1+\Delta _0)+1\right) \frac{P_{12}^2}{(x_{12}^2)^{2\Delta_0}}\ .
\end{align*}
}\noindent

\paragraph{$\boldsymbol{U(N)}$-case.} 
%%%%%%%%%%%%%%%%%%%%%%%%%%%%%%%%%%%%%%%%%%
The rest of the three-point correlation functions are for the $U(N)$-case
{\footnotesize{{\allowdisplaybreaks
\begin{align*}
    \langle J_0 J_0 [J_0J_0]_{0,0}\rangle &=2N(N+1) \Delta_0^2 \frac{1}{ (x_{13}^2)^{\Delta_0}(x_{23}^2)^{\Delta_0}}\ ,\nn\w2
    \langle J_0 J_0 [J_0J_0]_{0,2}\rangle &=2N\Delta_0^2 (1+\Delta_0)(1+2N(1+\Delta_0))\frac{Q_3^2}{ (x_{13}^2)^{\Delta_0}(x_{23}^2)^{\Delta_0}}\ ,
    \nn\w2
    \langle J_0 J_0 [J_0J_0]_{1,0}\rangle &= -2N(2N-1)\Delta_0^2\frac{1}{x_{12}^2 (x_{13}^2)^{\Delta_0+1}(x_{23}^2)^{\Delta_0+1}}\ ,
    \nn\w2
    \langle J_0 J_0 [J_0J_0]_{2,0}\rangle &=16N\Delta_0^2 (1+\Delta_0)(2+\Delta_0+2N(1+\Delta_0))\frac{1}{(x_{12}^2)^2 (x_{13}^2)^{\Delta_0+2}(x_{23}^2)^{\Delta_0+2}}\ ,\\
     \langle J_1 J_1 [J_0J_0]_{1,0}\rangle&=12 N(\Delta_0-1)^2 \Delta_0^2 \frac{(P_{12}+2 Q_1 Q_2)}{x_{12}^2(x_{13}^2)^{1+\Delta_0}(x_{23}^2)^{1+\Delta_0}} \ ,
    \nn \w2
    \langle J_1 J_1 [J_1J_1]_{1,0}\rangle&=4N (2 N+1) (\Delta_0-1)^4 \Delta_0^2 \frac{(P_{12}+2 Q_1 Q_2)}{x_{12}^2(x_{13}^2)^{1+\Delta_0}(x_{23}^2)^{1+\Delta_0}}\ ,
   \nn \w2
    \langle J_1 J_1 [J_1J_1]_{0,2}\rangle&=4N (2 N+1) (\Delta_0-1)^4 \Delta_0^2 \frac{P_{13}P_{23}}{(x_{13}^2)^{\Delta_0}(x_{23}^2)^{\Delta_0}}\ , 
    \nn\w2
    \langle J_1 J_1 [J_0J_0]_{0,2}\rangle&=-4N (\Delta_0-1)^2 \Delta_0^2 (\Delta_0+1) (2 \Delta_0+1) \frac{P_{13}P_{23}}{(x_{13}^2)^{1+\Delta_0}(x_{23}^2)^{1+\Delta_0}}\ , 
    \nn\w2
     \langle J_2 J_2 [J_0J_0]_{2,0}\rangle&=4N (2 N+1) (\Delta_0-1)^2 \Delta_0^2 (1+\Delta_0)(2+\Delta_0)(2+3\Delta_0)4+3\Delta_0)
     \frac{(P_{12}+2 Q_1 Q_2)}{x_{12}^2(x_{13}^2)^{2+\Delta_0}(x_{23}^2)^{2+\Delta_0}}\ , 
 \nn\w2
     \langle J_2 J_2 [J_2J_2]_{2,0}\rangle&=16N (\Delta_0-1)^4 \Delta_0^4 (1+\Delta_0)(2+\Delta_0)^2(\Delta_0+2N(1+\Delta_0))
     \frac{(P_{12}+2 Q_1 Q_2)}{x_{12}^2(x_{13}^2)^{2+\Delta_0}(x_{23}^2)^{2+\Delta_0}}\ , 
     \nn\w2
     \langle J_0 J_2 [J_0J_0]_{0,2}\rangle&=2N(\Delta_0-1)\Delta_0^2(1+\Delta_0)(2+\Delta_0)(1+2\Delta_0) \frac{P_{23}^2}{(x_{13}^2)^{\Delta_0}(x_{23}^2)^{\Delta_0}} \ ,
    \nn \w2
    \langle J_0 J_2 [J_0J_2]_{0,2}\rangle&=N(\Delta_0-1)^2\Delta_0^2(2+\Delta_0)(\Delta_0+2N(1+\Delta_0)) \frac{P_{23}^2}{(x_{13}^2)^{\Delta_0}(x_{23}^2)^{\Delta_0}} \ ,
    \nn\w2
    \langle J_1 J_1 [J_0J_0]_{2,0}\rangle&=64N (\Delta_0-1)^2 \Delta_0^2 (\Delta_0+1) (\Delta_0+2) (3 \Delta_0+2) (3 \Delta_0+4) 
    \frac{(P_{12}+2 Q_1 Q_2)^2}{(x_{12}^2)^2(x_{13}^2)^{2+\Delta_0}(x_{23}^2)^{2+\Delta_0}} \ ,
\end{align*}
\begin{align*}
    \langle [J_0J_0]_{0,0}[J_0J_0]_{0,0}\rangle&= 2N (1+N) \frac{1}{(x_{12}^2)^{2\Delta_0}}\ , \\
     \langle [J_0J_0]_{1,0}[J_0J_0]_{1,0}\rangle&= 12N (2 N-1) \Delta _0^2 \left(\Delta _0+2\right)  \frac{1}{(x_{12}^2)^{2(1+\Delta_0)}}\ , \\
     \langle [J_0J_0]_{2,0}[J_0J_0]_{2,0}\rangle&=16N\left(\Delta _0+1\right){}^2 \left(\Delta _0+4\right) \left(3 \Delta _0+2\right) \left(3 \Delta _0+4\right) \left(\Delta _0+4 \left(\Delta _0+1\right) N+2\right) \frac{1}{(x_{12}^2)^{2(2+\Delta_0)}}\ , \\
     \langle [J_0J_0]_{0,2}[J_0J_0]_{0,2}\rangle&=4N\Delta _0^2 \left(\Delta _0+1\right){}^2 \left(2 \Delta _0+1\right) \left(2 \left(\Delta _0+1\right) N+1\right) \frac{P_{12}^2}{(x_{12}^2)^{2\Delta_0}}\ , \\
    \langle [J_1J_1]_{0,2}[J_1J_1]_{0,2}\rangle&=4 N (2 N+1)\left(\Delta _0-1\right){}^4 \Delta _0^2  \frac{P_{12}^2}{(x_{12}^2)^{2\Delta_0}}\ , \\
    \langle [J_1J_1]_{1,0}[J_1J_1]_{1,0}\rangle&= 4 \left(\Delta _0-1\right){}^4 \Delta _0^2 \left(\Delta _0+2\right) N (2 N+1) \frac{1}{(x_{12}^2)^{2(1+\Delta_0)}}\ , \\
    \langle [J_1J_1]_{2,0}[J_1J_1]_{2,0}\rangle&=\frac{32 N^2\left(\Delta _0-1\right){}^4 \Delta _0^4 \left(\Delta _0+1\right){}^2 \left(\Delta _0+2\right){}^2 \left(\Delta _0+4\right) }{3 \Delta _0+4} \frac{1}{(x_{12}^2)^{2(2+\Delta_0)}}+\mathcal{O}(N)\ , \\
    \langle [J_0J_2]_{0,2}[J_0J_2]_{0,2}\rangle&=\left(\Delta _0-1\right){}^2 \Delta _0^2 \left(\Delta _0+2\right) N \left(\Delta _0+2 \left(\Delta _0+1\right) N\right) \frac{P_{12}^2}{(x_{12}^2)^{2\Delta_0}}\ .
\end{align*}
}}}\noindent

\paragraph{$\boldsymbol{USp(N)}$-case.}
%%%%%%%%%%%%%%%%%%%%%%%%%%%%%%%%%%%%%%%%%%%%%%
The relevant correlators are collected below
{\allowdisplaybreaks\footnotesize
\begin{align*}
    \langle J_0 J_0 [J_0J_0]_{1,0}\rangle&=-32 N(2N-1) \Delta_0^2 \frac{1}{x_{12}^2(x_{13}^2)^{1+\Delta_0}(x_{23}^2)^{1+\Delta_0}} \ ,
    \w2
    \langle J_1^{a_1b_1} J_1^{a_2b_2} [J_0J_0]_{1,0}\rangle&=48 N\Delta_0^2(\Delta_0-1)^2  \frac{(P_{12}+2 Q_1 Q_2)(\epsilon^{a_1b_2}\epsilon^{a_2b_1}-\epsilon^{a_1a_2}\epsilon^{b_1b_2})}{x_{12}^2(x_{13}^2)^{1+\Delta_0}(x_{23}^2)^{1+\Delta_0}} \ ,
    \w2  
    \langle J_1^{a_1b_1} J_1^{a_2b_2} [J_1J_1]_{1,0}^{a_2b_3,a_4b_4}\rangle&=4N^2\Delta_0^2(\Delta_0-1)^4\frac{(P_{12}+2 Q_1 Q_2)(\epsilon^{a_1a_3}\epsilon^{a_2a_4}\epsilon^{b_1b_3}\epsilon^{b_2 b_4}+\text{7 more})}{(x_{12}^2)^{2(\Delta_0+1)}} +\mathcal{O}(N)\ ,
    \w2
     \langle  [J_0J_0]_{0,0}  [J_0J_0]_{0,0}\rangle &=32 N(1+N) \frac{1}{(x_{12}^2)^{2\Delta_0}} \ ,
     \w2
    \langle  [J_0J_0]_{1,0}  [J_0J_0]_{1,0}\rangle &=192 N(2N-1) \Delta_0^2(2+\Delta_0) \frac{1}{(x_{12}^2)^{2(\Delta_0+1)}} \ ,
    \w2
    \langle [J_1J_1]_{1,0}^{a_1b_1,a_2b_2} [J_1J_1]_{1,0}^{a_2b_3,a_4b_4}\rangle&=4N^2\Delta_0^2(2+\Delta_0)(\Delta_0-1)^4\frac{1}{(x_{12}^2)^{2(\Delta_0+1)}}\ .
\end{align*}
}

%%%%%%%%%%%%%%%%%%%%%%%%%%%%%%%%%%%%%%%%%%%%%%%%%%%%%%%%%%
\end{appendix}
%%%%%%%%%%%%%%%%%%%%%%%%%%%%%%%%%%%%%%%%%%%%%%%%%%%%%%%%%%

%%%%%%%%%%%%%%%%%%%%%
\providecommand{\href}[2]{#2}\begingroup\raggedright\endgroup


\begin{thebibliography}{10}

\bibitem{Sundborg:2000wp}
B.~Sundborg, {\it {Stringy gravity, interacting tensionless strings and
  massless higher spins}},  {\em Nucl. Phys. Proc. Suppl.} {\bf 102} (2001)
  113--119 [\href{http://arXiv.org/abs/hep-th/0103247}{{\tt hep-th/0103247}}].
%%CITATION = HEP-TH/0103247;%%

\bibitem{Sezgin:2002rt}
E.~Sezgin and P.~Sundell, {\it {Massless higher spins and holography}},  {\em
  Nucl. Phys.} {\bf B644} (2002) 303--370
  [\href{http://arXiv.org/abs/hep-th/0205131}{{\tt hep-th/0205131}}]. [Erratum:
  Nucl. Phys.B660,403(2003)].
%%CITATION = HEP-TH/0205131;%%

\bibitem{Klebanov:2002ja}
I.~R. Klebanov and A.~M. Polyakov, {\it {AdS dual of the critical O(N) vector
  model}},  {\em Phys. Lett.} {\bf B550} (2002) 213--219
  [\href{http://arXiv.org/abs/hep-th/0210114}{{\tt hep-th/0210114}}].
%%CITATION = HEP-TH/0210114;%%

\bibitem{Sezgin:2003pt}
E.~Sezgin and P.~Sundell, {\it {Holography in 4D (super) higher spin theories
  and a test via cubic scalar couplings}},  {\em JHEP} {\bf 0507} (2005) 044
  [\href{http://arXiv.org/abs/hep-th/0305040}{{\tt hep-th/0305040}}].
%%CITATION = HEP-TH/0305040;%%

\bibitem{Gaberdiel:2010pz}
M.~R. Gaberdiel and R.~Gopakumar, {\it {An $AdS_3$ Dual for Minimal Model
  CFTs}},  {\em Phys.Rev.} {\bf D83} (2011) 066007
  [\href{http://arXiv.org/abs/1011.2986}{{\tt 1011.2986}}].
%%CITATION = ARXIV:1011.2986;%%

\bibitem{Bekaert:2015tva}
X.~Bekaert, J.~Erdmenger, D.~Ponomarev and C.~Sleight, {\it {Quartic AdS
  Interactions in Higher-Spin Gravity from Conformal Field Theory}},  {\em
  JHEP} {\bf 11} (2015) 149 [\href{http://arXiv.org/abs/1508.04292}{{\tt
  1508.04292}}].
%%CITATION = ARXIV:1508.04292;%%

\bibitem{Sleight:2017pcz}
C.~Sleight and M.~Taronna, {\it {Higher-Spin Gauge Theories and Bulk
  Locality}},  {\em Phys. Rev. Lett.} {\bf 121} (2018), no.~17 171604
  [\href{http://arXiv.org/abs/1704.07859}{{\tt 1704.07859}}].
%%CITATION = ARXIV:1704.07859;%%

\bibitem{Ponomarev:2017qab}
D.~Ponomarev, {\it {A Note on (Non)-Locality in Holographic Higher Spin
  Theories}},  {\em Universe} {\bf 4} (2018), no.~1 2
  [\href{http://arXiv.org/abs/1710.00403}{{\tt 1710.00403}}].
%%CITATION = ARXIV:1710.00403;%%

\bibitem{Blencowe:1988gj}
M.~Blencowe, {\it {A Consistent Interacting Massless Higher Spin Field Theory
  in $D$ = (2+1)}},  {\em Class.Quant.Grav.} {\bf 6} (1989) 443.
%%CITATION = CQGRD,6,443;%%

\bibitem{Campoleoni:2010zq}
A.~Campoleoni, S.~Fredenhagen, S.~Pfenninger and S.~Theisen, {\it {Asymptotic
  symmetries of three-dimensional gravity coupled to higher-spin fields}},
  {\em JHEP} {\bf 1011} (2010) 007 [\href{http://arXiv.org/abs/1008.4744}{{\tt
  1008.4744}}].
%%CITATION = ARXIV:1008.4744;%%

\bibitem{Henneaux:2010xg}
M.~Henneaux and S.-J. Rey, {\it {Nonlinear $W_{infinity}$ as Asymptotic
  Symmetry of Three-Dimensional Higher Spin Anti-de Sitter Gravity}},  {\em
  JHEP} {\bf 1012} (2010) 007 [\href{http://arXiv.org/abs/1008.4579}{{\tt
  1008.4579}}].
%%CITATION = ARXIV:1008.4579;%%

\bibitem{Tseytlin:2002gz}
A.~A. Tseytlin, {\it {On limits of superstring in AdS(5) x S**5}},  {\em Theor.
  Math. Phys.} {\bf 133} (2002) 1376--1389
  [\href{http://arXiv.org/abs/hep-th/0201112}{{\tt hep-th/0201112}}]. [Teor.
  Mat. Fiz.133,69(2002)].
%%CITATION = HEP-TH/0201112;%%

\bibitem{Segal:2002gd}
A.~Y. Segal, {\it {Conformal higher spin theory}},  {\em Nucl. Phys.} {\bf
  B664} (2003) 59--130 [\href{http://arXiv.org/abs/hep-th/0207212}{{\tt
  hep-th/0207212}}].
%%CITATION = HEP-TH/0207212;%%

\bibitem{Bekaert:2010ky}
X.~Bekaert, E.~Joung and J.~Mourad, {\it {Effective action in a higher-spin
  background}},  {\em JHEP} {\bf 02} (2011) 048
  [\href{http://arXiv.org/abs/1012.2103}{{\tt 1012.2103}}].
%%CITATION = ARXIV:1012.2103;%%

\bibitem{Ponomarev:2016lrm}
D.~Ponomarev and E.~D. Skvortsov, {\it {Light-Front Higher-Spin Theories in
  Flat Space}},  {\em J. Phys.} {\bf A50} (2017), no.~9 095401
  [\href{http://arXiv.org/abs/1609.04655}{{\tt 1609.04655}}].
%%CITATION = ARXIV:1609.04655;%%

\bibitem{Ponomarev:2017nrr}
D.~Ponomarev, {\it {Chiral Higher Spin Theories and Self-Duality}},  {\em JHEP}
  {\bf 12} (2017) 141 [\href{http://arXiv.org/abs/1710.00270}{{\tt
  1710.00270}}].
%%CITATION = ARXIV:1710.00270;%%

\bibitem{Skvortsov:2018jea}
E.~D. Skvortsov, T.~Tran and M.~Tsulaia, {\it {Quantum Chiral Higher Spin
  Gravity}},  {\em Phys. Rev. Lett.} {\bf 121} (2018), no.~3 031601
  [\href{http://arXiv.org/abs/1805.00048}{{\tt 1805.00048}}].
%%CITATION = ARXIV:1805.00048;%%

\bibitem{Skvortsov:2018uru}
E.~Skvortsov, {\it {Light-Front Bootstrap for Chern-Simons Matter Theories}},
  \href{http://arXiv.org/abs/1811.12333}{{\tt 1811.12333}}.
%%CITATION = ARXIV:1811.12333;%%

\bibitem{Joung:2015eny}
E.~Joung, S.~Nakach and A.~A. Tseytlin, {\it {Scalar scattering via conformal
  higher spin exchange}},  {\em JHEP} {\bf 02} (2016) 125
  [\href{http://arXiv.org/abs/1512.08896}{{\tt 1512.08896}}].
%%CITATION = ARXIV:1512.08896;%%

\bibitem{Ponomarev:2016jqk}
D.~Ponomarev and A.~A. Tseytlin, {\it {On quantum corrections in higher-spin
  theory in flat space}},  {\em JHEP} {\bf 05} (2016) 184
  [\href{http://arXiv.org/abs/1603.06273}{{\tt 1603.06273}}].
%%CITATION = ARXIV:1603.06273;%%

\bibitem{Beccaria:2016syk}
M.~Beccaria, S.~Nakach and A.~A. Tseytlin, {\it {On triviality of S-matrix in
  conformal higher spin theory}},  {\em JHEP} {\bf 09} (2016) 034
  [\href{http://arXiv.org/abs/1607.06379}{{\tt 1607.06379}}].
%%CITATION = ARXIV:1607.06379;%%

\bibitem{Hikida:2017ehf}
Y.~Hikida and T.~Uetoko, {\it {Correlators in higher-spin $AdS_3$ holography
  from Wilson lines with loop corrections}},  {\em PTEP} {\bf 2017} (2017)
  113B03 [\href{http://arXiv.org/abs/1708.08657}{{\tt 1708.08657}}].
%%CITATION = ARXIV:1708.08657;%%

\bibitem{Bekaert:2014cea}
X.~Bekaert, J.~Erdmenger, D.~Ponomarev and C.~Sleight, {\it {Towards
  holographic higher-spin interactions: Four-point functions and higher-spin
  exchange}},  {\em JHEP} {\bf 03} (2015) 170
  [\href{http://arXiv.org/abs/1412.0016}{{\tt 1412.0016}}].
%%CITATION = ARXIV:1412.0016;%%

\bibitem{Sleight:2016dba}
C.~Sleight and M.~Taronna, {\it {Higher Spin Interactions from Conformal Field
  Theory: The Complete Cubic Couplings}},  {\em Phys. Rev. Lett.} {\bf 116}
  (2016), no.~18 181602 [\href{http://arXiv.org/abs/1603.00022}{{\tt
  1603.00022}}].
%%CITATION = ARXIV:1603.00022;%%

\bibitem{Giombi:2013fka}
S.~Giombi and I.~R. Klebanov, {\it {One Loop Tests of Higher Spin AdS/CFT}},
  {\em JHEP} {\bf 12} (2013) 068 [\href{http://arXiv.org/abs/1308.2337}{{\tt
  1308.2337}}].
%%CITATION = ARXIV:1308.2337;%%

\bibitem{Giombi:2014iua}
S.~Giombi, I.~R. Klebanov and B.~R. Safdi, {\it {Higher Spin
  AdS$_{d+1}$/CFT$_d$ at One Loop}},  {\em Phys. Rev.} {\bf D89} (2014), no.~8
  084004 [\href{http://arXiv.org/abs/1401.0825}{{\tt 1401.0825}}].
%%CITATION = ARXIV:1401.0825;%%

\bibitem{Giombi:2014yra}
S.~Giombi, I.~R. Klebanov and A.~A. Tseytlin, {\it {Partition Functions and
  Casimir Energies in Higher Spin $AdS_{d+1}/CFT_d$}},  {\em Phys. Rev.} {\bf
  D90} (2014), no.~2 024048 [\href{http://arXiv.org/abs/1402.5396}{{\tt
  1402.5396}}].
%%CITATION = ARXIV:1402.5396;%%

\bibitem{Beccaria:2014xda}
M.~Beccaria and A.~A. Tseytlin, {\it {Higher spins in AdS$_{5}$ at one loop:
  vacuum energy, boundary conformal anomalies and AdS/CFT}},  {\em JHEP} {\bf
  11} (2014) 114 [\href{http://arXiv.org/abs/1410.3273}{{\tt 1410.3273}}].
%%CITATION = ARXIV:1410.3273;%%

\bibitem{Beccaria:2014jxa}
M.~Beccaria, X.~Bekaert and A.~A. Tseytlin, {\it {Partition function of free
  conformal higher spin theory}},  {\em JHEP} {\bf 08} (2014) 113
  [\href{http://arXiv.org/abs/1406.3542}{{\tt 1406.3542}}].
%%CITATION = ARXIV:1406.3542;%%

\bibitem{Beccaria:2014zma}
M.~Beccaria and A.~A. Tseytlin, {\it {Vectorial AdS$_5$/CFT$_4$ duality for
  spin-one boundary theory}},  {\em J. Phys.} {\bf A47} (2014), no.~49 492001
  [\href{http://arXiv.org/abs/1410.4457}{{\tt 1410.4457}}].
%%CITATION = ARXIV:1410.4457;%%

\bibitem{Beccaria:2014qea}
M.~Beccaria, G.~Macorini and A.~A. Tseytlin, {\it {Supergravity one-loop
  corrections on AdS$_7$ and AdS$_3$, higher spins and AdS/CFT}},  {\em Nucl.
  Phys.} {\bf B892} (2015) 211--238 [\href{http://arXiv.org/abs/1412.0489}{{\tt
  1412.0489}}].
%%CITATION = ARXIV:1412.0489;%%

\bibitem{Basile:2014wua}
T.~Basile, X.~Bekaert and N.~Boulanger, {\it {Flato-Fronsdal theorem for
  higher-order singletons}},  {\em JHEP} {\bf 11} (2014) 131
  [\href{http://arXiv.org/abs/1410.7668}{{\tt 1410.7668}}].
%%CITATION = ARXIV:1410.7668;%%

\bibitem{Beccaria:2015vaa}
M.~Beccaria and A.~A. Tseytlin, {\it {On higher spin partition functions}},
  {\em J. Phys.} {\bf A48} (2015), no.~27 275401
  [\href{http://arXiv.org/abs/1503.08143}{{\tt 1503.08143}}].
%%CITATION = ARXIV:1503.08143;%%

\bibitem{Beccaria:2016tqy}
M.~Beccaria and A.~A. Tseytlin, {\it {Iterating free-field AdS/CFT: higher spin
  partition function relations}},  {\em J. Phys.} {\bf A49} (2016), no.~29
  295401 [\href{http://arXiv.org/abs/1602.00948}{{\tt 1602.00948}}].
%%CITATION = ARXIV:1602.00948;%%

\bibitem{Bae:2016rgm}
J.-B. Bae, E.~Joung and S.~Lal, {\it {One-loop test of free SU(N ) adjoint
  model holography}},  {\em JHEP} {\bf 04} (2016) 061
  [\href{http://arXiv.org/abs/1603.05387}{{\tt 1603.05387}}].
%%CITATION = ARXIV:1603.05387;%%

\bibitem{Bae:2016hfy}
J.-B. Bae, E.~Joung and S.~Lal, {\it {On the Holography of Free Yang-Mills}},
  {\em JHEP} {\bf 10} (2016) 074 [\href{http://arXiv.org/abs/1607.07651}{{\tt
  1607.07651}}].
%%CITATION = ARXIV:1607.07651;%%

\bibitem{Pang:2016ofv}
Y.~Pang, E.~Sezgin and Y.~Zhu, {\it {One Loop Tests of Supersymmetric Higher
  Spin $AdS_4/CFT_3$}},  \href{http://arXiv.org/abs/1608.07298}{{\tt
  1608.07298}}.
%%CITATION = ARXIV:1608.07298;%%

\bibitem{Gunaydin:2016amv}
M.~G{\"u}naydin, E.~D. Skvortsov and T.~Tran, {\it {Exceptional $F(4)$
  higher-spin theory in AdS$_{6}$ at one-loop and other tests of duality}},
  {\em JHEP} {\bf 11} (2016) 168 [\href{http://arXiv.org/abs/1608.07582}{{\tt
  1608.07582}}].
%%CITATION = ARXIV:1608.07582;%%

\bibitem{Giombi:2016pvg}
S.~Giombi, I.~R. Klebanov and Z.~M. Tan, {\it {The ABC of Higher-Spin
  AdS/CFT}},  \href{http://arXiv.org/abs/1608.07611}{{\tt 1608.07611}}.
%%CITATION = ARXIV:1608.07611;%%

\bibitem{Bae:2016xmv}
J.-B. Bae, E.~Joung and S.~Lal, {\it {A note on vectorial AdS$_{5}$/CFT$_{4}$
  duality for spin-$j$ boundary theory}},  {\em JHEP} {\bf 12} (2016) 077
  [\href{http://arXiv.org/abs/1611.00112}{{\tt 1611.00112}}].
%%CITATION = ARXIV:1611.00112;%%

\bibitem{Brust:2016xif}
C.~Brust and K.~Hinterbichler, {\it {Partially Massless Higher-Spin Theory II:
  One-Loop Effective Actions}},  {\em JHEP} {\bf 01} (2017) 126
  [\href{http://arXiv.org/abs/1610.08522}{{\tt 1610.08522}}].
%%CITATION = ARXIV:1610.08522;%%

\bibitem{Skvortsov:2017ldz}
E.~D. Skvortsov and T.~Tran, {\it {AdS/CFT in Fractional Dimension and Higher
  Spin Gravity at One Loop}},  {\em Universe} {\bf 3} (2017), no.~3 61
  [\href{http://arXiv.org/abs/1707.00758}{{\tt 1707.00758}}].
%%CITATION = ARXIV:1707.00758;%%

\bibitem{Vasiliev:1992av}
M.~A. Vasiliev, {\it More on equations of motion for interacting massless
  fields of all spins in (3+1)-dimensions},  {\em Phys. Lett.} {\bf B285}
  (1992) 225--234.
%%CITATION = PHLTA,B285,225;%%

\bibitem{Giombi:2009wh}
S.~Giombi and X.~Yin, {\it {Higher Spin Gauge Theory and Holography: The
  Three-Point Functions}},  {\em JHEP} {\bf 1009} (2010) 115
  [\href{http://arXiv.org/abs/0912.3462}{{\tt 0912.3462}}].
%%CITATION = ARXIV:0912.3462;%%

\bibitem{Giombi:2010vg}
S.~Giombi and X.~Yin, {\it {Higher Spins in AdS and Twistorial Holography}},
  {\em JHEP} {\bf 1104} (2011) 086 [\href{http://arXiv.org/abs/1004.3736}{{\tt
  1004.3736}}].
%%CITATION = ARXIV:1004.3736;%%

\bibitem{Boulanger:2015ova}
N.~Boulanger, P.~Kessel, E.~D. Skvortsov and M.~Taronna, {\it {Higher spin
  interactions in four-dimensions: Vasiliev versus Fronsdal}},  {\em J. Phys.}
  {\bf A49} (2016), no.~9 095402 [\href{http://arXiv.org/abs/1508.04139}{{\tt
  1508.04139}}].
%%CITATION = ARXIV:1508.04139;%%

\bibitem{Skvortsov:2015lja}
E.~D. Skvortsov and M.~Taronna, {\it {On Locality, Holography and Unfolding}},
  {\em JHEP} {\bf 11} (2015) 044 [\href{http://arXiv.org/abs/1508.04764}{{\tt
  1508.04764}}].
%%CITATION = ARXIV:1508.04764;%%

\bibitem{Vasiliev:2016xui}
M.~A. Vasiliev, {\it {Current Interactions and Holography from the 0-Form
  Sector of Nonlinear Higher-Spin Equations}},  {\em JHEP} {\bf 10} (2017) 111
  [\href{http://arXiv.org/abs/1605.02662}{{\tt 1605.02662}}].
%%CITATION = ARXIV:1605.02662;%%

\bibitem{Gelfond:2017wrh}
O.~A. Gelfond and M.~A. Vasiliev, {\it {Current Interactions from the One-Form
  Sector of Nonlinear Higher-Spin Equations}},  {\em Nucl. Phys.} {\bf B931}
  (2018) 383--417 [\href{http://arXiv.org/abs/1706.03718}{{\tt 1706.03718}}].
%%CITATION = ARXIV:1706.03718;%%

\bibitem{Misuna:2017bjb}
N.~Misuna, {\it {On current contribution to Fronsdal equations}},  {\em Phys.
  Lett.} {\bf B778} (2018) 71--78 [\href{http://arXiv.org/abs/1706.04605}{{\tt
  1706.04605}}].
%%CITATION = ARXIV:1706.04605;%%

\bibitem{Didenko:2018fgx}
V.~E. Didenko, O.~A. Gelfond, A.~V. Korybut and M.~A. Vasiliev, {\it {Homotopy
  Properties and Lower-Order Vertices in Higher-Spin Equations}},  {\em J.
  Phys.} {\bf A51} (2018), no.~46 465202
  [\href{http://arXiv.org/abs/1807.00001}{{\tt 1807.00001}}].
%%CITATION = ARXIV:1807.00001;%%

\bibitem{Kazinski:2005eb}
P.~O. Kazinski, S.~L. Lyakhovich and A.~A. Sharapov, {\it {Lagrange structure
  and quantization}},  {\em JHEP} {\bf 07} (2005) 076
  [\href{http://arXiv.org/abs/hep-th/0506093}{{\tt hep-th/0506093}}].
%%CITATION = HEP-TH/0506093;%%

\bibitem{Boulanger:2015kfa}
N.~Boulanger, E.~Sezgin and P.~Sundell, {\it {4D Higher Spin Gravity with
  Dynamical Two-Form as a Frobenius-Chern-Simons Gauge Theory}},
  \href{http://arXiv.org/abs/1505.04957}{{\tt 1505.04957}}.
%%CITATION = ARXIV:1505.04957;%%

\bibitem{Petkou:2003zz}
A.~C. Petkou, {\it {Evaluating the AdS dual of the critical O(N) vector
  model}},  {\em JHEP} {\bf 03} (2003) 049
  [\href{http://arXiv.org/abs/hep-th/0302063}{{\tt hep-th/0302063}}].
%%CITATION = HEP-TH/0302063;%%

\bibitem{Koch:2010cy}
R.~de~Mello~Koch, A.~Jevicki, K.~Jin and J.~P. Rodrigues, {\it {$AdS_4/CFT_3$
  Construction from Collective Fields}},  {\em Phys. Rev.} {\bf D83} (2011)
  025006 [\href{http://arXiv.org/abs/1008.0633}{{\tt 1008.0633}}].
%%CITATION = ARXIV:1008.0633;%%

\bibitem{Penedones:2010ue}
J.~Penedones, {\it {Writing CFT correlation functions as AdS scattering
  amplitudes}},  {\em JHEP} {\bf 03} (2011) 025
  [\href{http://arXiv.org/abs/1011.1485}{{\tt 1011.1485}}].
%%CITATION = ARXIV:1011.1485;%%

\bibitem{Fitzpatrick:2011hu}
A.~L. Fitzpatrick and J.~Kaplan, {\it {Analyticity and the Holographic
  S-Matrix}},  {\em JHEP} {\bf 10} (2012) 127
  [\href{http://arXiv.org/abs/1111.6972}{{\tt 1111.6972}}].
%%CITATION = ARXIV:1111.6972;%%

\bibitem{Fitzpatrick:2011dm}
A.~L. Fitzpatrick and J.~Kaplan, {\it {Unitarity and the Holographic
  S-Matrix}},  {\em JHEP} {\bf 10} (2012) 032
  [\href{http://arXiv.org/abs/1112.4845}{{\tt 1112.4845}}].
%%CITATION = ARXIV:1112.4845;%%

\bibitem{Cardona:2017tsw}
C.~Cardona, {\it {Mellin-(Schwinger) representation of One-loop Witten diagrams
  in AdS}},  \href{http://arXiv.org/abs/1708.06339}{{\tt 1708.06339}}.
%%CITATION = ARXIV:1708.06339;%%

\bibitem{Giombi:2017hpr}
S.~Giombi, C.~Sleight and M.~Taronna, {\it {Spinning AdS Loop Diagrams: Two
  Point Functions}},  \href{http://arXiv.org/abs/1708.08404}{{\tt 1708.08404}}.
%%CITATION = ARXIV:1708.08404;%%

\bibitem{Yuan:2017vgp}
E.~Y. Yuan, {\it {Loops in the Bulk}},
  \href{http://arXiv.org/abs/1710.01361}{{\tt 1710.01361}}.
%%CITATION = ARXIV:1710.01361;%%

\bibitem{Bertan:2018khc}
I.~Bertan and I.~Sachs, {\it {Loops in Anti–de Sitter Space}},  {\em Phys.
  Rev. Lett.} {\bf 121} (2018), no.~10 101601
  [\href{http://arXiv.org/abs/1804.01880}{{\tt 1804.01880}}].
%%CITATION = ARXIV:1804.01880;%%

\bibitem{Bertan:2018afl}
I.~Bertan, I.~Sachs and E.~D. Skvortsov, {\it {Quantum $\phi^4$ Theory in
  AdS${}_4$ and its CFT Dual}},  {\em JHEP} {\bf 02} (2019) 099
  [\href{http://arXiv.org/abs/1810.00907}{{\tt 1810.00907}}].
%%CITATION = ARXIV:1810.00907;%%

\bibitem{Alday:2016njk}
L.~F. Alday, {\it {Large Spin Perturbation Theory for Conformal Field
  Theories}},  {\em Phys. Rev. Lett.} {\bf 119} (2017), no.~11 111601
  [\href{http://arXiv.org/abs/1611.01500}{{\tt 1611.01500}}].
%%CITATION = ARXIV:1611.01500;%%

\bibitem{Aharony:2016dwx}
O.~Aharony, L.~F. Alday, A.~Bissi and E.~Perlmutter, {\it {Loops in AdS from
  Conformal Field Theory}},  {\em JHEP} {\bf 07} (2017) 036
  [\href{http://arXiv.org/abs/1612.03891}{{\tt 1612.03891}}].
%%CITATION = ARXIV:1612.03891;%%

\bibitem{Caron-Huot:2017vep}
S.~Caron-Huot, {\it {Analyticity in Spin in Conformal Theories}},  {\em JHEP}
  {\bf 09} (2017) 078 [\href{http://arXiv.org/abs/1703.00278}{{\tt
  1703.00278}}].
%%CITATION = ARXIV:1703.00278;%%

\bibitem{Alday:2017xua}
L.~F. Alday and A.~Bissi, {\it {Loop Corrections to Supergravity on $AdS_5
  \times S^5$}},  {\em Phys. Rev. Lett.} {\bf 119} (2017), no.~17 171601
  [\href{http://arXiv.org/abs/1706.02388}{{\tt 1706.02388}}].
%%CITATION = ARXIV:1706.02388;%%

\bibitem{Aprile:2017bgs}
F.~Aprile, J.~M. Drummond, P.~Heslop and H.~Paul, {\it {Quantum Gravity from
  Conformal Field Theory}},  {\em JHEP} {\bf 01} (2018) 035
  [\href{http://arXiv.org/abs/1706.02822}{{\tt 1706.02822}}].
%%CITATION = ARXIV:1706.02822;%%

\bibitem{Aprile:2017xsp}
F.~Aprile, J.~M. Drummond, P.~Heslop and H.~Paul, {\it {Unmixing
  Supergravity}},  {\em JHEP} {\bf 02} (2018) 133
  [\href{http://arXiv.org/abs/1706.08456}{{\tt 1706.08456}}].
%%CITATION = ARXIV:1706.08456;%%

\bibitem{Alday:2017vkk}
L.~F. Alday and S.~Caron-Huot, {\it {Gravitational S-matrix from CFT dispersion
  relations}},  {\em JHEP} {\bf 12} (2018) 017
  [\href{http://arXiv.org/abs/1711.02031}{{\tt 1711.02031}}].
%%CITATION = ARXIV:1711.02031;%%

\bibitem{Aprile:2017qoy}
F.~Aprile, J.~M. Drummond, P.~Heslop and H.~Paul, {\it {Loop corrections for
  Kaluza-Klein AdS amplitudes}},  {\em JHEP} {\bf 05} (2018) 056
  [\href{http://arXiv.org/abs/1711.03903}{{\tt 1711.03903}}].
%%CITATION = ARXIV:1711.03903;%%

\bibitem{Alday:2018pdi}
L.~F. Alday, A.~Bissi and E.~Perlmutter, {\it {Genus-One String Amplitudes from
  Conformal Field Theory}},  \href{http://arXiv.org/abs/1809.10670}{{\tt
  1809.10670}}.
%%CITATION = ARXIV:1809.10670;%%

\bibitem{Ghosh:2018bgd}
K.~Ghosh, {\it {Polyakov-Mellin Bootstrap for AdS loops}},
  \href{http://arXiv.org/abs/1811.00504}{{\tt 1811.00504}}.
%%CITATION = ARXIV:1811.00504;%%

\bibitem{Maldacena:2011jn}
J.~Maldacena and A.~Zhiboedov, {\it {Constraining Conformal Field Theories with
  A Higher Spin Symmetry}},  \href{http://arXiv.org/abs/1112.1016}{{\tt
  1112.1016}}.
%%CITATION = ARXIV:1112.1016;%%

\bibitem{Boulanger:2013zza}
N.~Boulanger, D.~Ponomarev, E.~D. Skvortsov and M.~Taronna, {\it {On the
  uniqueness of higher-spin symmetries in AdS and CFT}},  {\em Int. J. Mod.
  Phys.} {\bf A28} (2013) 1350162 [\href{http://arXiv.org/abs/1305.5180}{{\tt
  1305.5180}}].
%%CITATION = ARXIV:1305.5180;%%

\bibitem{Alba:2013yda}
V.~Alba and K.~Diab, {\it {Constraining conformal field theories with a higher
  spin symmetry in d=4}},  \href{http://arXiv.org/abs/1307.8092}{{\tt
  1307.8092}}.
%%CITATION = ARXIV:1307.8092;%%

\bibitem{Alba:2015upa}
V.~Alba and K.~Diab, {\it {Constraining conformal field theories with a higher
  spin symmetry in $d> 3$ dimensions}},
  \href{http://arXiv.org/abs/1510.02535}{{\tt 1510.02535}}.
%%CITATION = ARXIV:1510.02535;%%

\bibitem{Saadi:1989tb}
M.~Saadi and B.~Zwiebach, {\it {Closed String Field Theory from Polyhedra}},
  {\em Annals Phys.} {\bf 192} (1989) 213.
%%CITATION = APNYA,192,213;%%

\bibitem{Mack:1976pa}
G.~Mack, {\it {Convergence of Operator Product Expansions on the Vacuum in
  Conformal Invariant Quantum Field Theory}},  {\em Commun. Math. Phys.} {\bf
  53} (1977) 155.
%%CITATION = CMPHA,53,155;%%

\bibitem{Sotkov:1976xe}
G.~M. Sotkov and R.~P. Zaikov, {\it {Conformal Invariant Two Point and Three
  Point Functions for Fields with Arbitrary Spin}},  {\em Rept. Math. Phys.}
  {\bf 12} (1977) 375.
%%CITATION = RMHPB,12,375;%%

\bibitem{Osborn:1993cr}
H.~Osborn and A.~C. Petkou, {\it {Implications of conformal invariance in field
  theories for general dimensions}},  {\em Annals Phys.} {\bf 231} (1994)
  311--362 [\href{http://arXiv.org/abs/hep-th/9307010}{{\tt hep-th/9307010}}].
%%CITATION = HEP-TH/9307010;%%

\bibitem{Costa:2011mg}
M.~S. Costa, J.~Penedones, D.~Poland and S.~Rychkov, {\it {Spinning Conformal
  Correlators}},  {\em JHEP} {\bf 11} (2011) 071
  [\href{http://arXiv.org/abs/1107.3554}{{\tt 1107.3554}}].
%%CITATION = ARXIV:1107.3554;%%

\bibitem{Ponomarev:2019ofr} 
  D.~Ponomarev, {\it {From bulk loops to boundary large-N expansion}},
    \href{http://arXiv.org/abs/1908.03974}{{\tt 1908.03974}}.
  %%CITATION = ARXIV:1908.03974;%%

\bibitem{Taronna:2016ats}
M.~Taronna, {\it {Pseudo-local Theories: A Functional Class Proposal}},  in
  {\em {Proceedings, International Workshop on Higher Spin Gauge Theories:
  Singapore, Singapore, November 4-6, 2015}}, pp.~59--84, 2017.
\newblock \href{http://arXiv.org/abs/1602.08566}{{\tt 1602.08566}}.
%%CITATION = ARXIV:1602.08566;%%

\bibitem{Bekaert:2016ezc}
X.~Bekaert, J.~Erdmenger, D.~Ponomarev and C.~Sleight, {\it {Bulk quartic
  vertices from boundary four-point correlators}},  in {\em {Proceedings,
  International Workshop on Higher Spin Gauge Theories: Singapore, Singapore,
  November 4-6, 2015}}, pp.~291--303, 2017.
\newblock \href{http://arXiv.org/abs/1602.08570}{{\tt 1602.08570}}.
%%CITATION = ARXIV:1602.08570;%%

\bibitem{Rastelli:2017udc}
L.~Rastelli and X.~Zhou, {\it {How to Succeed at Holographic Correlators
  Without Really Trying}},  {\em JHEP} {\bf 04} (2018) 014
  [\href{http://arXiv.org/abs/1710.05923}{{\tt 1710.05923}}].
%%CITATION = ARXIV:1710.05923;%%

\bibitem{Pius:2016jsl}
R.~Pius and A.~Sen, {\it {Cutkosky rules for superstring field theory}},  {\em
  JHEP} {\bf 10} (2016) 024 [\href{http://arXiv.org/abs/1604.01783}{{\tt
  1604.01783}}]. [Erratum: JHEP09,122(2018)].
%%CITATION = ARXIV:1604.01783;%%

\bibitem{Giddings:1999qu}
S.~B. Giddings, {\it {The Boundary S matrix and the AdS to CFT dictionary}},
  {\em Phys. Rev. Lett.} {\bf 83} (1999) 2707--2710
  [\href{http://arXiv.org/abs/hep-th/9903048}{{\tt hep-th/9903048}}].
%%CITATION = HEP-TH/9903048;%%

\bibitem{Dusedau:1985ue}
D.~W. Dusedau and D.~Z. Freedman, {\it {Lehmann Spectral Representation for
  Anti-de Sitter Quantum Field Theory}},  {\em Phys. Rev.} {\bf D33} (1986)
  389.
%%CITATION = PHRVA,D33,389;%%

\bibitem{Beisert:2004di}
N.~Beisert, M.~Bianchi, J.~F. Morales and H.~Samtleben, {\it {Higher spin
  symmetry and N=4 SYM}},  {\em JHEP} {\bf 07} (2004) 058
  [\href{http://arXiv.org/abs/hep-th/0405057}{{\tt hep-th/0405057}}].
%%CITATION = HEP-TH/0405057;%%

\bibitem{Craigie:1983fb}
N.~S. Craigie, V.~K. Dobrev and I.~T. Todorov, {\it {Conformally Covariant
  Composite Operators in Quantum Chromodynamics}},  {\em Annals Phys.} {\bf
  159} (1985) 411--444.
%%CITATION = APNYA,159,411;%%

\bibitem{Sleight:2018epi}
C.~Sleight and M.~Taronna, {\it {Spinning Mellin Bootstrap: Conformal Partial
  Waves, Crossing Kernels and Applications}},  {\em Fortsch. Phys.} {\bf 66}
  (2018), no.~8-9 1800038 [\href{http://arXiv.org/abs/1804.09334}{{\tt
  1804.09334}}].
%%CITATION = ARXIV:1804.09334;%%

\end{thebibliography}
\end{document}